\documentclass[12pt, draftclsnofoot, onecolumn]{IEEEtran}
\usepackage{epsfig,latexsym}
\usepackage{float}
\usepackage{indentfirst}
\usepackage{amsmath}
\usepackage{amssymb}
\usepackage{times}
\usepackage{subfigure}
\usepackage{psfrag}
\usepackage{hyperref}
\usepackage{cite}
\usepackage{lastpage}
\usepackage{fancyhdr}
\usepackage{color}
\usepackage{amsthm}
\usepackage{bigints}
\usepackage{bm}
\usepackage{tabularx}
\usepackage{booktabs}
\usepackage{stfloats}
\newtheorem{Lemma}{Lemma}
\newtheorem{lemma}[Lemma]{$\mathbf{Lemma}$}
\newtheorem{Proposition}{Proposition}
\newtheorem{proposition}[Proposition]{$\mathbf{Proposition}$}
\newtheorem{Corollary}{Corollary}
\newtheorem{corollary}[Corollary]{$\mathbf{Corollary}$}

\usepackage{epsfig} 
\usepackage{amsmath} 
\usepackage{amssymb} 
\usepackage{bm} 
\usepackage{tabularx} 
\usepackage{booktabs} 
\usepackage{algorithm,algorithmic} 

\def\*#1{\mathbf{#1}} 
\def\ei{\mathrm{Ei}} 
\def\e{\mathrm{e}}
\def\E{\mathrm{E}}
\def\pr{\mathrm{Pr}}
\def\st{\text{s.t.}}
\makeatletter 
\newcommand{\Rmnum}[1]{\expandafter\@slowromancap\romannumeral #1@} 
\makeatother 

\newcommand{\absS}[1]{\left|{#1}\right|^2}
\newcommand{\prb}[1]{\pr\left\{{#1}\right\}}
\newcommand{\fai}[2]{\varphi\left(#1,#2\right)}
\newcommand{\faip}[2]{\varphi'\left(#1,#2\right)}

\newcommand{\EI}[1]{\ei\left(#1\right)}
\newcommand{\LN}[1]{\ln\left(#1\right)}

\newcounter{remarkCnt}
\newcolumntype{Y}{>{\centering\arraybackslash}X}

\begin{document}

\title{Position Information-based NOMA for Downlink and Uplink Transmission in Mobile Scenarios}
\author{Hao Qiu, Shaoshuai Gao, GuoFang Tu, and Shouxin Zong \thanks{The authors are with the School of Electronic, Electrical and Communication Engineering, the University of Chinese Academy of Sciences, Beijing 101408, China.}\vspace{-2em}}

\maketitle

\begin{abstract}
In this paper, a non-orthogonal multiple access (NOMA) system with partial 
channel state information (CSI) for downlink and uplink transmission in mobile scenarios 
is considered, i.e., users are deployed randomly and will move casually around the base station (BS). 
In this case, the channel gain of each user varies over time,
which has an influence on the performance of conventional NOMA. 
An analytical framework is developed to evaluate 
the impact of position estimation deviation in terms of decoding order error probability, 
average sum rate and outage probability. 
Based on the framework, dynamic power allocation (DPA) for downlink NOMA 
and dynamic power control (DPC) for uplink NOMA
are put forward to optimize the outage performance with user distance information.
It has been shown that the performance of NOMA relies on accurate user position information.
To this end, two algorithms based on position filtering are proposed to improve the accuracy of user position.
Monte Carlo simulation is presented to demonstrate the improvement of spectrum efficiency and outage performance. Simulation results verify the accuracy of the proposed analytical framework.
\end{abstract}

\section{Introduction}
Non-orthogonal multiple access (NOMA) has recently received considerable attention
and has been recognized as a promising candidate for future wireless networks \cite{RN348}.
Compared with orthogonal multiple access (OMA), it is considered not only
to improve the spectral efficiency but also to support massive connectivity.
More users than the number of available orthogonal resource blocks can be supported \cite{RN353}. 
The key idea of NOMA is to realize multiple access (MA) 
by encouraging non-orthogonal resource allocation among users. 
In particular, NOMA in power domain (PD-NOMA) is inspired by the superposition coding (SC) technology, 
in which the signals of different users are multiplexed with different power levels. 
At receivers, successive interference cancellation (SIC) is implemented to decode
the superimposed messages \cite{RN350}.

NOMA can be applied into both downlink and uplink transmission. 
For downlink NOMA, the users with poor channel conditions are usually allocated more power
so that their signals are decoded by regarding the signals of others as noise.
In \cite{RN374}, the impact of path loss on the performance of NOMA with randomly deployed users 
was evaluated, which demonstrated that NOMA can outperform conventional OMA.
Some works investigated different types of resource allocation,
such as user pairing \cite{RN422} and power allocation \cite{RN399}.
For uplink NOMA, the power control scheme needs to be well-designed so that
the base station (BS) is capable of detecting signals from all users.
The uplink ergodic sum-rate gain of over OMA was shown in \cite{wei2020}.
In \cite{zhang2016}, a back-off power control scheme was proposed and 
its outage performance was analyzed. 
The joint user grouping and power control issue was considered in \cite{yang2016,ali2016}.

The aforementioned literature is under the assumption that 
both the BS and users can obtain perfect channel information. 
Actually, this assumption might not be realistic due to many limitations
like high mobility of users and channel estimation error. 
Therefore, some works focused on the case of partial channel state information(CSI) 
for downlink\cite{Arzykulov2019, li2020, liu2020, ALWANI2019} 
and uplink transmission\cite{gao2018,Schiessl2020}.
The impact of channel estimation error on NOMA system was investigated in \cite{Arzykulov2019,li2020}.
In \cite{liu2020}, statistical CSI was exploited to optimize the transmit beamforming vectors
and the power splitting ratio in the simultaneous wireless information and power transfer 
(SWIPT)-enabled cooperative NOMA system.
In \cite{ALWANI2019}, a novel beamforming and cluster strategy was proposed 
with quantized channel direction information.
Moreover, the impact of imperfect CSI on uplink NOMA detection was analyzed in \cite{gao2018}.
The authors of \cite{Schiessl2020} evaluated NOMA in a low-latency system
and showed that imperfect CSI creates a larger penalty for NOMA than for OMA.
For the past several years, some works focused on NOMA with distance information \cite{RN358}.
For instance, distance-based user selection can be adopted in NOMA systems, e.g., \cite{RN357,RN440}.
The performance of random beamforming NOMA network with distance information was analyzed in \cite{RN396}.
In \cite{RN500}, the authors verified the rationality of ranking users with distance information
for Rayleigh and Nakagami-m fading channels.

There has been massive research on NOMA performance evaluation, 
and much of them is with the aid of stochastic geometry tools\cite{StoGeo}. 
For instance, the number of users is fixed 
and they are assumed uniformly distributed within an area, such as \cite{RN374, RN358}. 
Poisson point process (PPP) or Poisson cluster process (PCP) are also frequently used 
to model the distribution of transceivers \cite{RN364, RN440, RN500, RN396, RN357}.
These articles focused on the static scenarios where users are deployed randomly 
and the locations are independent in different time slots. 
Recently, the authors of \cite{marshoud2016} applied NOMA to a visible light communication (VLC) system 
and the performance was evaluated by using the random walk model.
In \cite{liu2019}, a machine learning scheme was proposed to demonstrate 
the NOMA-enabled unmanned aerial vehicles (UAV) placement issue, 
in which users are roaming on the ground.
However, the research of NOMA in mobile scenarios is still limited 
and more research contribution is required. 

Different from previous static NOMA systems like \cite{RN358}, 
this paper considers a distance-based NOMA system in mobile scenarios, 
i.e., the users are deployed randomly around the BS and will move casually.
It is also different from our previous work \cite{8761437}
since uplink transmission will be considered in this work 
and a more comprehensive analysis on the effect of position estimation will be presented,
including outage performance, user pairing and path loss exponents.
Moreover, based on our derived analytical results, two optimization problems are formulated 
to improve the outage performance in downlink and uplink scenarios.
It is noticed that some existing works have studied power optimization problems with distance information,
such as power allocation with outage constraints \cite{cui2016}, 
outage balancing \cite{shi2016}, ergodic capacity maximization \cite{sun2015} 
and sum throughput maximization \cite{wang2018}. 
In this article, the common outage probability (COP) is considered 
to evaluate the system outage performance.
The COP describes the outage performance of the entire NOMA network rather than a single user,
which emphasizes the user fairness \cite{xu2016}.
To the best of our knowledge, this COP minimization with user distance information
has not been considered,
and we fill this gap by proposing two dynamic power schemes.
The main contributions of this article are summarized as follows:

\begin{itemize}
	\item We investigate the impact of position estimation error in the distance-based NOMA system.
	To this end, we adopt a new metric named decoding order error probability
	to evaluate the effect on user order.
	The analytical expressions for the average sum rate and outage probability are derived
	in the two-node pairing case\footnote{
	The results of more users can be extended with the similar method, 
	whereas the process becomes more complex. According to 
	the rationality analysis of NOMA with distance information in \cite{RN500}, $M=2$ or user pairing 
	is a more practical setting.}.
	For obtaining more insights, the closed-form approximation of average sum rate
	at high signal-to-noise ratios (SNR) is provided
	and indicates that the high-SNR slope is not limited by position estimation error.
	However, the outage analysis of uplink NOMA shows that an error floor exists
	when transmission power is fixed. 

	\item In order to further improve the outage performance and break the error floor of uplink NOMA,
	we propose a dynamic power allocation (DPA) scheme for downlink NOMA and 
	a dynamic power control (DPC) scheme for uplink NOMA. 
	More specifically, we optimize the signal power with distance information 
	from the perspective of COP minimization.

	\item We propose two algorithms based on position filtering 
	to increase the accuracy of position information,
	which are beneficial to improve the spectrum efficiency and outage performance.
	In the first algorithm (position tracking-based NOMA),
	the BS applies a Kalman Filter\cite{RN419} to track the movement of each user
	for obtaining more accurate position information.
	The second algorithm (position prediction-based NOMA) exploits the obtained 
	information of user mobility to predict their positions.
	It can be applied in the case where frequent position information feedback is unavailable,
	which strikes a good balance between system performance and network overhead.

\end{itemize} 

The reminder of this paper is organized as follows. 
Section \ref{s2} describes the system model.
In Section \ref{s3}, the impact of position estimation deviation on NOMA average sum rate 
and outage performance is evaluated, and dynamic signal power schemes are proposed. 
In Section \ref{s4}, the position filtering-based NOMA system is designed. 
Section \ref{s5} provides simulation results and some discussion. 
Finally, Section \ref{s6} concludes this paper.

\textbf{Notations:}
Matrices and vectors are denoted by upper- and lower-case boldface letters, respectively.
The superscript $T$ represents transpose.
Expectation is expressed by $\E\left\{\cdot \right\}$ 
and probability is described by $\pr\left\{\cdot \right\}$.
$\mathcal{N}\left(\*{a},\*{R}\right)$ and $\mathcal{CN}\left(\*{a},\*{R}\right)$ 
denote the distribution of real Gaussian random vectors 
and circularly symmetric complex Gaussian (CSCG) random vectors 
with mean vector $\*{a}$, covariance matrix $\*{R}$, respectively.
$\*{I}_n$ is the identity matrix of size $n$.
$\binom n k$ denotes the binomial coefficient.

\section{system model}\label{s2}

\begin{figure}[!htbp]\centering
\includegraphics[width=0.9\textwidth]{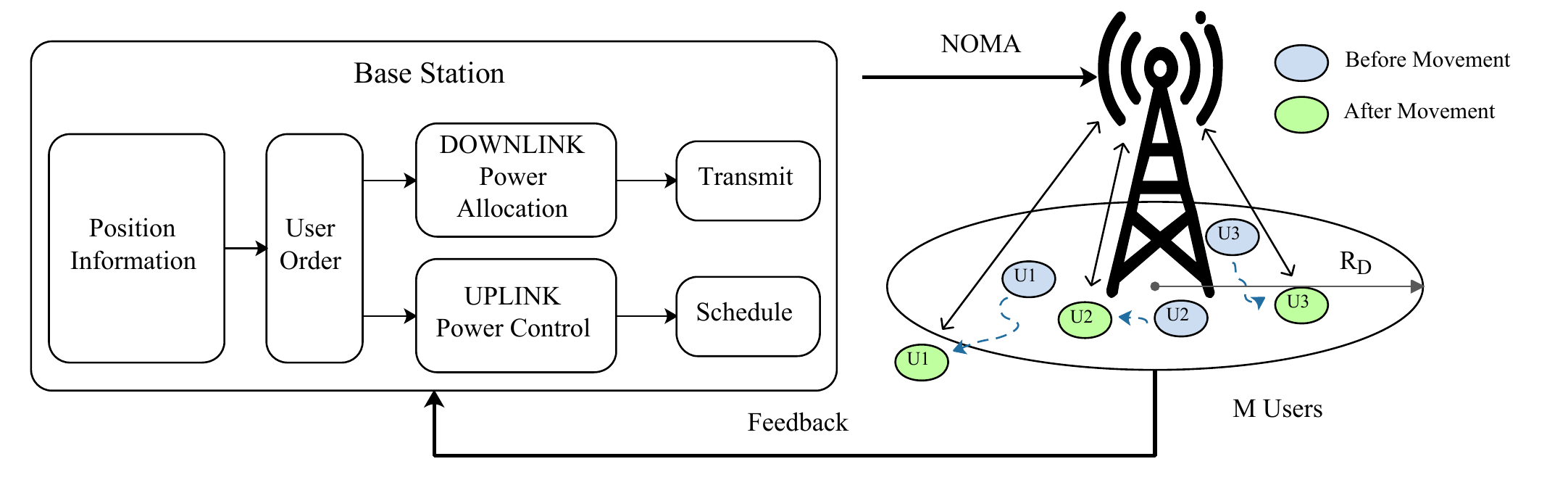}\vspace{-1em}
\caption{system model. \vspace{-1em}}
\label{system model}
\end{figure}

In this paper, we will focus on a single-cell NOMA network in a mobile scenario 
as shown in Fig. \ref{system model}.
The network consists of one BS and $M$ users $U_m\left(m=1,2,\ldots,M\right)$, 
where the BS and all the users are equipped with a single antenna.
It is assumed that the BS is located at the origin of a two-dimensional Euclidean plane, 
expressed as $\mathbb{R}^2$. 
At the beginning, the users are deployed randomly in the disc $D$ with radius $R_D$. 
After that, they are assumed to roam around the BS. 
The distance between the BS and $U_m$ is denoted by $d_m\left(m=1,2,\ldots,M\right)$.
For user position information, positioning of cellular networks can be utilized in the mobile scenario,
such as time of arrival (TOA) measurement methods or global positioning system (GPS) \cite{cellPos}.
The distance estimation between the BS and users is derived subsequently.
Then user order and pairing are determined according to the results as well.
We suppose that the position estimation deviation of axis-x and axis-y in the Cartesian coordinate
follows a Gaussian distribution with zero mean and variance $\sigma_{ob}^2$ for simplicity.
In general, the Gaussian distribution assumption is reasonable according to asymptotic arguments
and the central limit theorem \cite{RN418}.

On the other hand, wireless channels are assumed to be independent and quasi-static fading, 
i.e., the channel gain remains constant for a given coherence time 
and varies independently from one time slot to another. 
A composite channel model consisting of large-scale path loss 
and small-scale Rayleigh fading is considered.
Thus, the channel coefficient between user $U_m$ and the base station is denoted by 
$r_m=h_md_m^{-\alpha/2}$, 
where $h_m\sim\mathcal{CN}\left(0,1\right)$ and $\alpha$ is the path loss exponent. 
The channel gain is calculated as $\left|r_m\right|^2=\left|h_m\right|^2d_m^{-\alpha}$, 
in which $\left|h_m\right|^2$ follows an exponential distribution with parameter $1$.

\subsection{Downlink NOMA Scheme with Partial Channel Information}
In the downlink scenario, the BS transmits messages to the users simultaneously. 
Without loss of generality, their distances are ranked as $d_1<d_2<\ldots<d_M$.
The signals are superimposed with different power according to the principle of superposition coding. 
The signal sent by the BS is expressed as follows:
\begin{align}
S=\sum_{m=1}^M\sqrt{\alpha_mP}S_m,
\end{align}
where $S_m\left(m=1,2,\ldots,M\right)$ denotes the signal of the user $U_m$, 
$\alpha_m$ is its power allocation factor subject to $\sum_{m=1}^M\alpha_m=1$,
and $P$ denotes the transmission power.
At receivers, the signal received by user $U_m$ is formulated as
\begin{align}\label{rec sig}
y_m=r_m\sum_{l=1}^M\sqrt{\alpha_lP}S_l+n_m,
\end{align}
where $n_m$ denotes the Gaussian noise of user $U_m$. 
It is normalized with zero mean and variance $\sigma_n^2$, 
i.e., $n_m\sim\mathcal{CN}\left(0,\sigma^2_n\right)$. 
If SIC is carried out by user $U_m$, 
it has to detect the messages of $U_l\left(m+1\le l\le M\right)$ successfully 
before detecting its own message.
The achievable rate for $U_m$ to detect the signal $S_l$ is shown as follows:
\begin{align}\label{gen c}
R_{m\rightarrow l}=\log_2\left(1+\frac{\left|r_m\right|^2\alpha_l}{\left|r_m\right|^2a_{l-1}
	+\frac1\rho}\right),
\end{align}
where $a_{l-1}=\sum_{k=1}^{l-1}\alpha_k$, $a_0=0$, 
and $\rho=\frac P{\sigma_n^2}$ denotes the transmission SNR.

Note that for two users $U_i$ and $U_j$, we have the following property 
if $\left|r_i\right|^2>\left|r_j\right|^2$:
\begin{align} 
R_{i\rightarrow j}=&\log_2\left(1+\frac{\left|r_i\right|^2\alpha_j}{\left|r_i\right|^2a_{j-1}
	+\frac1\rho}\right) \nonumber \\
>&\log_2\left(1+\frac{\left|r_j\right|^2\alpha_j}{\left|r_j\right|^2a_{j-1}+\frac1\rho}\right)=R_j.
\end{align}
In other words, the achievable rates are restricted by the users with 
the poorest channel condition\footnote{
As a special case, 
the achievable rate of a signal is restricted by themselves only if they are ranked by perfect CSI,
such as \cite{RN374,RN422}. However, under the assumption of partial CSI, the property 
$\left|r_i\right|^2<\left|r_j\right|^2$ cannot be guaranteed despite of $d_i<d_j$.}.
Moreover, in the delay-tolerant communication, 
data rates will be determined according to real-time channel conditions
and the average sum rate is used as a performance criterion. 
Assuming that user $U_m$ always decodes the signals $S_l\left(m+1\le l\le M\right)$ correctly, 
the sum rate of the system is expressed as follows:
\begin{align}\label{sr def}
R_{sum}^{\Rmnum{1}}=&\sum_{l=1}^M\min_{1\le m\le l}\log_2\left(1+\frac{\left|r_m\right|^2\alpha_l}
	{\left|r_m\right|^2a_{l-1}+\frac1\rho}\right).
\end{align}

In delay-sensitive communication, messages will be transmitted at a fixed target transmission rate. 
A typical performance metric is outage probability, 
which is defined as the probability that the achievable rate under variable channel conditions 
is less than the target rate.
An outage event occurs at $U_m$ when $S_l\left(l\ge m\right)$ fails to be decoded.
By denoting the target rate as $R^*_0$\footnote{
Even though different target rate setting could improve the throughput of the system, 
the same rate $R_0$ for the users is assumed in order to guarantee the fairness.}, 
the outage probability of $U_m$ is expressed as
\begin{align}
P_{om}^{\Rmnum{1}} = 1- \pr\left\{\bigcap_{l=m}^{M}R_{m\rightarrow l}>R^*_0\right\}.
\end{align}

We consider the common outage probability (COP) of the network. 
An outage event occurs if any of the users is in outage. 
Therefore, the COP is given as follows:
\begin{align}
P_{cop}^{\Rmnum{1}} = 1-\pr\left\{\bigcap_{m=1}^M\bigcap_{l=m}^{M}R_{m\rightarrow l}>R^*_0\right\}.
\label{cop def}
\end{align}

Some works are based on the assumption that fixed power allocation factors 
are predetermined no matter which decoding order is selected.
Actually, this assumption could lead to poor user fairness 
or a substantial amount of power is wasted to guarantee the QoS of poor users. 
To tackle this issue, we will allocate the power to users according to the SIC detection order. 

\subsection{Uplink NOMA Scheme with Partial Channel Information}
In uplink scenarios, users send signals to the base station at the same time, 
and they can also be ranked according to distance like downlink.
By denoting the transmission power of $U_m$ by $P_m$, 
the signal received by the BS is formulated as:
\begin{align}
y = \sum_{m=1}^M r_m\sqrt{P_m}S_m+ n,
\end{align}
where $n$ denotes the Gaussian noise. 
The BS detects the messages by a SIC detector. Unlike downlink, 
the decoding process starts from the nearest user to the farthest user to make sure that 
all the signals can be decoded.
The achievable rate of the signal $S_m$ is expressed like this:
\begin{align}
R_m = \log_2\left(1+\frac{\absS{r_m}P_m}{\sum_{l=m+1}^{M}\absS{r_l}P_l+\sigma_n^2}\right).
\end{align}

Similar to \eqref{sr def} and \eqref{cop def}, 
the sum rate and outage probability of uplink NOMA are formulated in the following:
\begin{align}
&R_{sum}^{\Rmnum{2}}=\sum_{m=1}^{M}R_m=\log_2\left(1+\sum_{m=1}^{M}\left|r_m\right|^2\rho_m\right) 
\label{up SR}, \\
&P_{cop}^{\Rmnum{2}}=1-\prb{\bigcap_{m=1}^{M}R_m>R_0^*},
\end{align}
where $\rho_m$ is the transmission SNR of $U_m$, i.e. $\rho_m=P_m/\sigma_n^2$.

It is worthy to point out that the transmission power of each user is constrained individually.
This can be equivalently expressed by largest transmission SNR $\Omega_m$,
which means $\rho_m<\Omega_m$.

Note that these metrics are obtained in the case that
the optimal channel coding and modulation scheme is adopted. 
The system considering practical coding and modulation schemes is 
beyond the scope of this paper, which is likely a promising future direction.

\section{performance analysis and power optimization}\label{s3}
In this section, the effect of the position estimation on the system performance is considered.  
We consider the case when two users are paired for implementation of NOMA like \cite{RN422}. 
Without loss of generality, it is assumed that $U_1$ and $U_2$ are paired and $d_1<d_2$.
The estimated distance of user $U_m$ is denoted by $\hat{d}_m$.

\subsection{decoding order error probability}
In the distance-based NOMA system, user order may be suboptimal 
due to channel fading and position estimation error.
We adopt a new metric to evaluate their effects on user order, 
which is named decoding order error probability. 
It is defined as the probability that the estimated distance-based order
is not in accordance with the CSI-based order\footnote{
Accuracy probability was defined in \cite{RN500}, 
which means the event when distance-based ranking yields the same results 
as instantaneous channel information. The decoding order error probability can be regarded 
as the complementary accuracy probability considering position estimation error. 
The optimal decoding order has been investigated in \cite{RN399} and \cite{gao2018}.}, 
which is expressed as
\begin{align}
P_e =& \prb{\hat{d}_1<\hat{d}_2,\absS{r_1}<\absS{r_2}} \nonumber \\
	 &+\prb{\hat{d}_1>\hat{d}_2,\absS{r_1}>\absS{r_2}}. \label{pe def}
\end{align}
This metric is used to derive the average sum rate and outage performance later in this section.

\begin{lemma}\label{LEMMA1}
In the fading-free scenario, the decoding order error probability of 
distance-based NOMA for the two pairing users with fixed locations 
is expressed as follows:
\begin{align}
P_e^1=& \prb{\hat{d}_1>\hat{d}_2} \nonumber \\
	=& \sum_{i=0}^{\infty}\sum_{j=0}^{\infty}P_{\lambda_1\beta}\left(i\right)
	P_{\lambda_2\beta}\left(j\right)I_{i,j}, \label{pe awgn}
\end{align}
where $I_{i,j}=\left(\frac12\right)^{\alpha_i+\alpha_j}\binom{\alpha_i+\alpha_j-1}{\alpha_j}
F\left(1,\alpha_i+\alpha_j,\alpha_j+1,\frac12\right)$,
$\alpha_i=i+1$,
$F\left(\cdot\right)$ denotes a hypergeometric function.
$P_\lambda\left(k\right)=\frac{e^{-\lambda}\lambda^k}{k!}\left(k\ge 0\right)$ 
is the probability mass function (PMF) of a Poisson distribution, $\lambda_k=d_k^2\left(k=1,2\right)$, 
$\beta=\frac1{2\sigma_{ob}^2}$.
\end{lemma}
\begin{IEEEproof}
Please refer to Appendix \ref{fading-free pe pr}.
\end{IEEEproof}

The lemma indicates that the decoding order error probability is expressed as a weighted sum.
The term $I_{i,j}$ is just the function of its indices and
the weight coefficients are the probabilities of Poisson distribution.
Note that \eqref{pe awgn} is convergent
because $P_e^1$ is an integral of a joint probability density function as shown in \eqref{pe ori}.
As a result, this probability $P_e^1$ can be approximated as a sum of finite terms. 

In fading-free scenarios, the user order error is only caused by position estimation deviation,
and $P_e^1$ is actually the probability of distance order error.
Furthermore, the effect of Rayleigh fading on decoding order error is considered in the following.

\begin{corollary}\label{corollary1}
For the Rayleigh fading channel, the decoding order error probability of 
distance-based NOMA for the two pairing users with fixed locations is expressed as follows:
\begin{align}\label{Pe ray}
P_e^{2}= \frac{D-1}{D+1}P_e^1+\frac1{D+1},
\end{align}
where $D={d_2^\alpha}/{d_1^\alpha}$ and the probability $P_e^1$ has been calculated 
in Lemma \ref{LEMMA1}.
\end{corollary}
\begin{IEEEproof}
We substitute \eqref{pe awgn} into \eqref{pe def}. 
Due to the independence of channel fading and position estimation, the probability is given by
\begin{align}
P_e^2=&\pr\left(\left|r_1\right|^2>\left|r_2\right|^2\right)\times P_e^1 \nonumber \\
&+\left[1-\pr\left(\left|r_1\right|^2>\left|r_2\right|^2\right)\right]\times\left(1-P_e^1\right). 
\label{pe2 ori}
\end{align}

Let $D={d_2^\alpha}/{d_1^\alpha}$. With some algebraic manipulations, 
the probability term in \eqref{pe2 ori} is calculated as
\begin{align}\label{pr fading}
\pr\left(\left|r_1\right|^2>\left|r_2\right|^2\right)=\pr\left(\left|h_2\right|^2<
	D\left|h_1\right|^2\right)=\frac{D}{D+1}.
\end{align}

By substituting \eqref{pr fading} into \eqref{pe2 ori}, this corollary is proved.
\end{IEEEproof}

Given $P_e^1<0.5$ and $D>1$, one can infer from \eqref{Pe ray} that
the error probability is larger than that of the fading-free scenario. 
The decoding order error event may still occur without positioning deviation due to incomplete CSI. 
Furthermore, \eqref{Pe ray} indicates that there is a linear relationship between $P_e^1$ and $P_e^2$.
And the impact of decoding order error on the system performance 
will be investigated with $P_e^1$ in the following.
The conclusions are the same with $P_e^2$ as well.

\subsection{downlink NOMA system}
In downlink transmission, the power is allocated to the two users.
Let $\beta$ denotes the larger power factor, and $1-\beta$ is for the other.

\begin{proposition}\label{PROSR1}
In downlink distance-based NOMA for two pairing users with fixed locations, 
the average sum rate performance is calculated as follows:
\begin{align}
R_{sum}^{\Rmnum{1}}= &\left(1-P_e^1\right)\faip{1}{\rho\left(1-\beta\right)}
	+P_e^1\faip{2}{\rho\left(1-\beta\right)} \nonumber \\
					 &+\fai{1}{\rho}+\fai{2}{\rho}-\fai{1}{\rho\left(1-\beta\right)} \nonumber \\
					 &-\fai{2}{\rho\left(1-\beta\right)}, \label{down sr eq}
\end{align}
where $\varphi\left(k,\phi\right)=-\frac{\lambda_k}{\lambda \ln2}\e^{\frac\lambda\phi}
{\mathrm{Ei}}\left(-\frac\lambda\phi\right)$ and 
$\varphi'\left(k,\phi\right)=-\frac1{\ln2}\e^{\frac{\lambda_k}{\phi}}
\ei\left(-\frac{\lambda_k}\phi\right)$, 
$\lambda_k=d_k^\alpha\left(k=1,2\right)$, $\lambda=\lambda_1+\lambda_2$,
$\ei\left(x\right)$ denotes an exponential integral.
The probability $P_e^1$ has been calculated in Lemma \ref{LEMMA1}.

In the high-SNR region, the average sum rate is approximated as
\begin{align} 
R_{sum}^{\Rmnum{1}}\approx \log_2\frac{\rho}{\lambda_1} - P_e^1\log_2\frac{\lambda_2}{\lambda_1}
	- \frac{C}{\ln2},
\end{align}
where $C$ denotes the Euler's constant \cite{RN416}.
\end{proposition}
\begin{IEEEproof}
Please refer to Appendix \ref{down sr pr}.
\end{IEEEproof}

Expressing the function $\varphi'\left(k,\phi\right)$ in the form of integral, 
we have $\varphi'\left(k,\phi\right)=\frac{\phi}{\ln2}\int^{\infty}_{0}
\frac{\e^{-\lambda_kt}}{1+t\phi}dt$. 
One can see that this function monotonically decreases with $\lambda_k$, 
which indicates $\varphi'\left(1,\rho\left(1-\beta\right)\right)>
\varphi'\left(2,\rho\left(1-\beta\right)\right)$ due to $\lambda_1<\lambda_2$.
Therefore, the decoding order error will lead to degradation of the NOMA spectrum efficiency
according to \eqref{down sr eq}.
The high-SNR approximation shows that the high-SNR slope \cite{li2020}
is not affected by the position deviation and capacity ceilings do not exist.

The optimal power allocation from the perspectives of average sum rate maximization
is achieved when all the power is assigned to the near user, 
i.e., $\beta=0$, which is also revealed in \cite{sun2015}.
Obviously, this strategy is unrealistic and 
the PA factor should be adjusted considering user fairness constraint.
For convenience of analysis, a fixed PA factor is considered in this paper
when the average sum rate is analyzed.

\begin{proposition} \label{PROPO1}
In downlink distance-based NOMA for two pairing users with fixed locations, 
let $\epsilon_0$ denotes the target SNR of the two signals, i.e., $\epsilon_0=2^{R_0^*}-1$. 
The COP is given as follows:
\begin{align}
P_{cop}^{\Rmnum{1}}=1-\left(1-P_e^1\right)\e^{-\left(\lambda_2A+\lambda_1\zeta\right)}
	-P_e^1\e^{-\left(\lambda_1A+\lambda_2\zeta\right)}, \label{down cop eq}
\end{align}
where $A=\frac{\epsilon_0}{\rho\left[\beta-\left(1-\beta\right)\epsilon_0\right]}$, 
$\beta-\left(1-\beta\right)\epsilon_0>0$, $B=\frac{\epsilon_0}{\rho\left(1-\beta\right)}$ 
$\zeta=\max\left\{A,B\right\}$, and $\lambda_k=d_k^\alpha\left(k=1,2\right)$.
The $P_e^1$ is decoding order error probability from Lemma \ref{LEMMA1}.
\end{proposition}
\begin{IEEEproof}
Please refer to Appendix \ref{down po pr}.
\end{IEEEproof}

An observation is that system parameter setting plays an important role 
in the outage performance and determines how the positioning deviation affects the COP. 
Firstly, $\beta>\left(1-\beta\right)\epsilon_0$ needs to be satisfied.
Otherwise, the outage probability will always be one.
On the condition $A>B$, i.e. $\beta<\left(1-\beta\right)\left(\epsilon_0+1\right)$, 
the outage probability is simplified as $P_{cop}^{\Rmnum{1}}=1-\e^{-\lambda A}$. 
This indicates that positioning deviation has no effect on the COP performance. 
If $A<B$, i.e. $\beta>\left(1-\beta\right)\left(\epsilon_0+1\right)$, 
positioning deviation will lead to the deterioration of the outage performance. 
When the difference between $A$ and $B$ is large or the difference between $\lambda_1$
and $\lambda_2$ is large, the impact of estimation deviation will become greater.

In order to improve the outage performance, 
we propose a dynamic power allocation (DPA) scheme for downlink NOMA based on our analysis above.
Our objective is to minimize the COP performance under the total power constraint. 
Note that the decoding order error probability is a performance metric 
to evaluate the effect of position estimation, which is unknown to the base station.
However, the power allocation can be optimized by using the estimated position information.
In this case, even though the approximation is suboptimal, 
the DPA scheme is practical and effective. 
We derive its closed-form solution and numerical results in Section \ref{s5} demonstrate the validity.
The problem is expressed mathematically as follows:
\begin{subequations}\label{problemDown}
\begin{align}
& \min_{\beta} ~ 1-\e^{-\left(\lambda_2A+\lambda_1\zeta\right)}, \\
& \st 		   ~ 0\le\beta\le1, \\
& ~~~~~~		 \beta>\left(1-\beta\right)\epsilon_0.
\end{align}
\end{subequations}

\begin{corollary}\label{OPTPA}
The optimal solution to the problem \eqref{problemDown} is given by
\begin{align} \label{solve_down}
\beta^*=\frac
{\sqrt{1+\epsilon_0}\left(\epsilon_0\lambda_1-\lambda_2\right)+\sqrt{\lambda_1\lambda_2}}
{\sqrt{1+\epsilon_0}\left[\lambda_1\left(1+\epsilon_0\right)-\lambda_2\right]}.
\end{align}
\end{corollary}
\begin{IEEEproof}
Please refer to Appendix \ref{problemDown pr}.
\end{IEEEproof}

\subsection{uplink NOMA system}
Different from the downlink NOMA, the average sum rate of uplink is always the same 
no matter which user is decoded first. 
The analytical result is given in the following.

\begin{proposition} \label{PROSR2}
In uplink distance-based NOMA for two pairing users with fixed locations, 
the average sum rate is calculated as follows:
\begin{align}
R_{sum}^{\Rmnum{2}} = &\left[-\frac{\lambda_1}{\rho_1}\varphi'\left(2,\rho_2\right)
	+\frac{\lambda_2}{\rho_2}\varphi'\left(1,\rho_1\right)\right] 
	\times \frac{1}{\frac{\lambda_2}{\rho_2}-\frac{\lambda_1}{\rho_1}}. \label{up sr eq}
\end{align}
In the high-SNR region, the approximate average sum rate is expressed as
\begin{align}
R_{sum}^{\Rmnum{2}}\approx \left(\frac{\lambda_1}{\rho_1}\log_2\frac{\lambda_2}{\rho_2}- 
	\frac{\lambda_2}{\rho_2}\log_2\frac{\lambda_1}{\rho_1}\right) \times 
	\frac{1}{\frac{\lambda_2}{\rho_2}-\frac{\lambda_1}{\rho_1}}- \frac{C}{\ln2}.
\end{align}
\end{proposition}
\begin{IEEEproof}
Refer to \eqref{up SR}, the average sum rate is written as follows:
\begin{align}
\E\left\{C\right\}=& \frac{1}{\rho_1\rho_2}\iint\log_2\left(1+x+y\right)
	\lambda_1\e^{-\frac{\lambda_1x}{\rho_1}}\lambda_2\e^{-\frac{\lambda_2y}{\rho_2}}dxdy.
\end{align}

Let $g=\left(\frac{\lambda_1}{\rho_1}+\frac{\lambda_2}{\rho_2}\right)/2$ 
and $h=\left(\frac{\lambda_2}{\rho_2}-\frac{\lambda_1}{\rho_1}\right)/2$. 
By replacing the integral variables by $t=x+y$ and $s=y-x$, we have
\begin{align}
\E\left\{C\right\}=& \frac{g^2-h^2}{2h}\int_0^{\infty}\log_2
	\left(1+t\right)\e^{-gt}\left(\e^{ht}-\e^{-ht}\right)dt. \label{ec1 mid}
\end{align}

After substituting \eqref{int res} and \eqref{fai2 def} into \eqref{ec1 mid}, 
the exact average sum rate is obtained. 
Moreover, the high-SNR approximation can be obtained 
by substituting \eqref{fai2 approx} into \eqref{up sr eq}.
\end{IEEEproof}

\begin{proposition} \label{PROPO2}
In uplink distance-based NOMA for two pairing users with fixed locations, 
the COP is expressed as
\begin{align}
P_{cop}^{\Rmnum{2}} =& 1-\left(1-P_e^1\right)\frac{\lambda_2}{\lambda_2+k\lambda_1}
	\e^{-\lambda_1C-\left(\lambda_2+k\lambda_1\right)B} \nonumber \\
					 &-P_e^1\times\frac{\lambda_1}{\lambda_1+k\lambda_2}
					 	\e^{-\lambda_2C-\left(\lambda_1+k\lambda_2\right)B}, \label{up po eq}
\end{align}
where $B=\frac{\epsilon_0}{\rho_2}$, $C=\frac{\epsilon_0}{\rho_1}$,
$k=\frac{\epsilon_0\rho_2}{\rho_1}$, and $\lambda_k=d_k^\alpha\left(k=1,2\right)$.
$P_e^1$ is calculated in Lemma \ref{LEMMA1}.
\end{proposition}
\begin{IEEEproof}
Please refer to Appendix \ref{up po pr}.
\end{IEEEproof}

An important observation is that when the transmission power approaches infinity,
i.e., $\rho_1\rightarrow\infty$ and $\rho_2\rightarrow\infty$,
the COP of uplink NOMA is given by
\begin{align}
P_{cop}^{\Rmnum{2}} = 1-\left(1-P_e^1\right)\frac{\lambda_2}{\lambda_2+k\lambda_1}
	-P_e^1\times\frac{\lambda_1}{\lambda_1+k\lambda_2}.
\end{align}
The result shows that the COP is determined by the power ratio of the two users in high SNR region. 
An error floor exists when transmission power ratio is fixed. 
This indicates the importance of user power control in uplink NOMA.
Based on our analysis, we put forward a dynamic power control scheme (DPC)
to break the error floor. 
After the base station obtains the estimated positions, an optimal transmission power 
to minimize the COP performance can be found under individual power constraint. 
Like the downlink PA optimization problem \eqref{problemDown}, the uplink PC optimization problem 
is formulated as follows:
\begin{subequations}\label{problemUp}
\begin{align}
& \min_{\rho_1, \rho_2} ~ 1-\frac{\lambda_2}{\lambda_2+k\lambda_1}
	\e^{-\lambda_1C-\left(\lambda_2+k\lambda_1\right)B}, \\
& \st 		   ~ 0\le\rho_1\le\Omega_1, \\
& ~~~~~~		 0\le\rho_2\le\Omega_2,
\end{align}
\end{subequations}
where $\Omega_k$ is the largest transmission power of user $k$. 

\begin{corollary}\label{OPTPC}
The optimal solution to the problem \eqref{problemUp} is given by
\begin{subequations} \label{solve_up}
\begin{gather}
\rho_1 = \Omega_1, \\
\rho_2 = \min\left(\Omega_2, \frac{\epsilon_0\lambda_1\lambda_2
	+\lambda_2\sqrt{4\Omega_1\lambda_1+\epsilon_0^2\lambda_1^2}}{2\lambda_1}\right).
\end{gather}
\end{subequations}
\end{corollary}
\begin{IEEEproof}
Please refer to Appendix \ref{problemUp pr}.
\end{IEEEproof}

\section{position filtering-based algorithms}\label{s4}
According to our previous analysis, 
the SIC strategy relies on accurate position information.
Thus in this section, we propose two position filtering-based algorithms, 
where the movement of users is traceable and this knowledge can be exploited 
to improve the system performance.

\subsection{User Mobile Model}
In this paper, the movement of users is described by a velocity sensor model \cite{RN418}.
This mobility model can be expressed by a continuous-time state-space model. 
The mobile state of a user is defined by a vector as follows:
\begin{align}
\*{s}\left(t\right)=\left[x\left(t\right), v_x\left(t\right), y\left(t\right), v_y\left(t\right)\right]^T,
\end{align}
where $x\left(t\right)$ and $y\left(t\right)$ specify the position 
in the Cartesian coordinate at time $t$. 
$v_x\left(t\right)$ and $v_y\left(t\right)$ denote the velocities in x-axis and y-axis, respectively. 
Thus, the distance between the user and the base station is calculated as
\begin{align}\label{pos2dis}
d^2\left(t\right)=x^2\left(t\right)+y^2\left(t\right).
\end{align}

Because the velocities change at any time, the variation can be expressed by a white noise 
stochastic process vector $\*{w}\left(t\right)$.
Its covariance matrix is denoted by $\tilde{\*{Q}}$.
\begin{align}
\*{w}\left(t\right)=\left[w_x\left(t\right),w_y\left(t\right)\right]^T, \qquad \tilde{\*{Q}}=\sigma_w^2\*{I}_2.
\end{align}
The state transition model is expressed with a linear differential equation as
\begin{align}
\dot{\*{s}}\left(t\right)=\tilde{\*{A}}\*s\left(t\right)+\tilde{\*{B}}\*w\left(t\right),
\end{align}
where $\tilde{\*{A}}$ and $\tilde{\*{B}}$ are state-transition matrix and random input matrix, 
respectively. They are presented as
\begin{align}
\tilde{\*{A}}=\left[
\begin{matrix} 
0 & 1 & 0 & 0 \\
0 & 0 & 0 & 0 \\
0 & 0 & 0 & 1 \\
0 & 0 & 0 & 0
\end{matrix}
\right], \qquad
\tilde{\*{B}}=\left[
\begin{matrix}
1 & 0 \\
0 & 0 \\
0 & 1 \\
0 & 0
\end{matrix}
\right].
\end{align}

At time $t$, a measurement of the state vector is made as
\begin{align}
\*{z}\left(t\right)=\*{Hs}\left(t\right)+\*{n}_{ob}\left(t\right),
\end{align}
where $\*{H}$ is the observation matrix and is presented as
\begin{align}
\*{H}=\left[
\begin{matrix}
1 & 0 & 0 & 0 \\
0 & 0 & 1 & 0
\end{matrix}
\right],
\end{align}
$\*{n}_{ob}\left(t\right)$ is the observation noise which is zero mean Gaussian white noise process with covariance matrix $\*{R}=\sigma_{ob}^2\*{I}_2$

After sampling the state and measurement vector every $T$ seconds, 
the model is transformed into discrete-time state-space model. 
Let $\*{s}_k=\*{s}\left(kT\right)$ and $\*{z}_k=\*{z}\left(kT\right)$, we have
\begin{align}
\*{s}_{k+1}=\*{As}_k+\*{w}_k,
\end{align}
where
\begin{align}
\*{A}=e^{\tilde{\*{A}}T}=\left[
\begin{matrix} 
1 & T & 0 & 0 \\
0 & 1 & 0 & 0 \\
0 & 0 & 1 & T \\
0 & 0 & 0 & 1
\end{matrix}
\right],
\end{align}
\begin{align}
\*{w}_k=\int_{kT}^{\left(k+1\right)T}e^{\tilde{\*{A}}\left(\left(k+1\right)T-\tau\right)}\tilde{\*{B}}\*{w}\left(\tau\right)d\tau,
\end{align}
where $\*{w}_k$ is a zero mean discrete-time Gaussian white noise vector, 
thus $\E\left\{\*{w}_k\*{w}_{k+i}^T\right\}=0$ for $i\ne0$. Besides, 
the covariance matrix $\*{Q}$ is calculated as
\begin{align}
\*{Q} = \E\left\{\*{w}_k\*{w}_k^T\right\}=\left[
\begin{matrix}
T\sigma_w^2 & 0 & 0 & 0 \\
0 & 0 & 0 & 0 \\
0 & 0 & T\sigma_w^2 & 0 \\
0 & 0 & 0 & 0
\end{matrix}
\right].
\end{align}

Based on this, we propose two algorithms based on position filtering in mobile scenarios.

\subsection{Position tracking-based NOMA}
The position tracking-based NOMA carries out position estimation with a Kalman filter,
which is based on the principle of minimizing error covariance \cite{RN419}. 
The algorithm is divided into two steps as follows:

{\it Step 1:} According to the user's state vector, covariance matrix and the observation data, 
Kalman filter algorithm is performed to obtain the optimal estimated position. 
Then the distance information is calculated by substituting the result into \eqref{pos2dis}.

{\it Step 2:} 
Users are ranked according to the estimated position and signal power is obtained according to 
predetermined power scheme.
Signals are superimposed at the transmitter for downlink NOMA or 
they are transmitted simultaneously for uplink NOMA.
At receivers, SIC decoding is carried out. 

In general, the Kalman filter can be divided into two phases, i.e., prediction and update.
In the prediction phase, the estimation of $\*{s}_k$ and its error covariance
are denoted by $\hat{\*{s}}_{k|k-1}$ and $\*{P}_{k|k-1}$, respectively.
They are denoted by $\hat{\*{s}}_{k|k}$ and $\*{P}_{k|k}$ in the update phase.
Let $\*{K}_k$ denotes the Kalman gain.
The detailed algorithm procedure is described in Algorithm \ref{algo1}.

\begin{figure}[H]
\begin{algorithm}[H]
\caption{Position Tracking-Based NOMA}
\label{algo1}
\begin{algorithmic}[1]
\REQUIRE State vector measurement $\*{z}_k$.
\ENSURE User order and signal power.
\\ \textbf{Initialization}
\STATE Set initial predicted state estimate $\hat{\*{s}}_{1|0}=\*{0}$ and initial predicted error covariance $\*{P}_{1|0}=\*{I}_4$.
\\ \textbf{Stage 1: Position filtering}
\FOR {$U_i \left(i=1...M\right)$}
	\STATE Calculate the Kalman gain: \\
	$\*{K}_k=\*{P}_{k|k-1}\*{H}^T\left(\*{HP}_{k|k-1}\*{H}^T+\*{R}\right)^{-1}$.
	\STATE Updated state equation: \\
	$\hat{\*{s}}_{k|k}=\hat{\*{s}}_{k|k-1}+\*{K}_k\left(\*{z}_k-\*{H}\hat{\*{s}}_{k|k-1}\right)$.
	\STATE Updated estimate covariance: \\
	$\*{P}_{k|k}=\left(\*{I}_4-\*{K}_k\*{H}\right)\*{P}_{k|k-1}$.
	\STATE Calculate the estimated distance: $\hat{d}_k=\hat{x}^2_k+\hat{y}_k^2$, \\
	where ($\hat{x}_k$, $\hat{y}_k$) is the position coordinate in $\hat{\*{s}}_{k|k}$.
	\STATE Predicted state equation: $\hat{\*{s}}_{k+1|k}=\*{A}\hat{\*{s}}_{k|k}$.
	\STATE Predicted error covariance: $\*{P}_{k+1|k}=\*{A}\*{P}_{k|k}\*{A}^T+\*{Q}$.
\ENDFOR
\\ \textbf{Stage 2: Superposition Coding}
\STATE Sort users $\hat{d}_1\le\hat{d}_2\le\ldots\le\hat{d}_M$.
\STATE User pairing and Scheduling. \\
\IF {Dynamic power scheme}
\STATE Downlink: Calculate PA factor according to \eqref{solve_down}. \\
Uplink: Calculate signal power according to \eqref{solve_up}.
\ELSIF {Fixed power scheme}
\STATE Downlink: larger PA factor is allocated to far user. \\
Uplink: Users transmit signals at maximum individual power constraint.
\ENDIF
\STATE Go to Step 2 for the next time slot $(k+1)$.
\end{algorithmic}
\end{algorithm}
\end{figure}

\subsection{Position prediction-based NOMA}
Now we consider a more realistic scenario. 
We know that transmitting the feedback information to the BS may result in 
a mount of communication overhead and extra transmission delay. 
In some cases, the position information may be unavailable for some slots.
Based on these, we propose the position prediction-based algorithm with less feedback information,
while the system performance can also be guaranteed.
In the process of position estimation, some knowledge about user mobility is obtained by the BS. 
This kind of knowledge can be used to predict users' position.

It should be noted that the prediction process is performed at each time slot 
whether with observation data or not,
while the state update process is not needed at the slots without enough observation data.
Then the NOMA algorithm is carried out in the same way as the position tracking-based NOMA.
The detailed algorithm procedure is shown in Algorithm \ref{algo2}.

\begin{figure}[H]
\begin{algorithm}[H]
\caption{Position Prediction-Based NOMA}
\label{algo2}
\begin{algorithmic}[1]
\REQUIRE State vector measurement $\*{z}_k$.
\ENSURE User order and signal power.
\\ \textbf{Initialization}
\STATE Set initial predicted state estimate $\hat{\*{s}}_{1|0}=\*{0}$ and initial predicted error covariance $\*{P}_{1|0}=\*{I}_4$.
\\ \textbf{Stage 1: Position filtering}
\FOR {$U_i \left(i=1...M\right)$}
\IF {The measurement is available}
\STATE Track the movement of the user the same way as Algorithm 1 (Step 3 to Step 8).
\ELSE
\STATE Calculate the estimated distance: $\hat{d}_k=\hat{x}^2_k+\hat{y}_k^2$, \\
where ($\hat{x}_k$, $\hat{y}_k$) is the position coordinate in $\hat{\*{s}}_{k|k-1}$.
\STATE Predicted state equation: \\
$\hat{\*{s}}_{k+1|k}=\*{A}\hat{\*{s}}_{k|k-1}$.
\STATE Predicted error covariance: \\
$\*{P}_{k+1|k}=\*{A}\*{P}_{k|k-1}\*{A}^T+\*{Q}$.
\ENDIF
\ENDFOR
\\ \textbf{Stage 2 is the same as that in Algorithm 1}
\end{algorithmic}
\end{algorithm}
\end{figure}

\section{numerical results and discussions}\label{s5}
\subsection{simulation results}

In this section, numerical results are provided to validate the correctness of our analytical results 
and evaluate the performance of the proposed NOMA schemes. 
The parameters used in our simulations are set as follows. 
The sampling interval is set to $T=0.2$ s and each sample trajectory includes 
$K=300$ sample points, which corresponds to the duration $1$ min. 
Three types of mobility model are considered to mimic the movement of users 
including random walk model, random waypoint model and Gauss-Markov model\cite{CampA}.
The mobile trajectory of each trial is independent.
The small-scale fading is assumed to be Rayleigh fading.
The thermal noise is set as $\sigma_n^2=-50$ dBm.
Monte Carlo simulation results are averaged over $10^6$ independent trials.

In terms of OMA benchmark, each user is allocated to one subchannel. 
The COP is calculated as follows:
\begin{align}
P_{om}^{\Rmnum{3}} = 1 - \e^{-\frac{\epsilon_0'\left(\lambda_1+\lambda_2\right)}{\rho}}, \label{oma po}
\end{align}
where $\rho$ denotes the transmission SNR and $\epsilon_0'$ is the target SNR in OMA, 
i.e., $\epsilon_0'=2^{2R_0^*}-1$.

\begin{figure}[H]\centering
\includegraphics[width=0.45\textwidth]{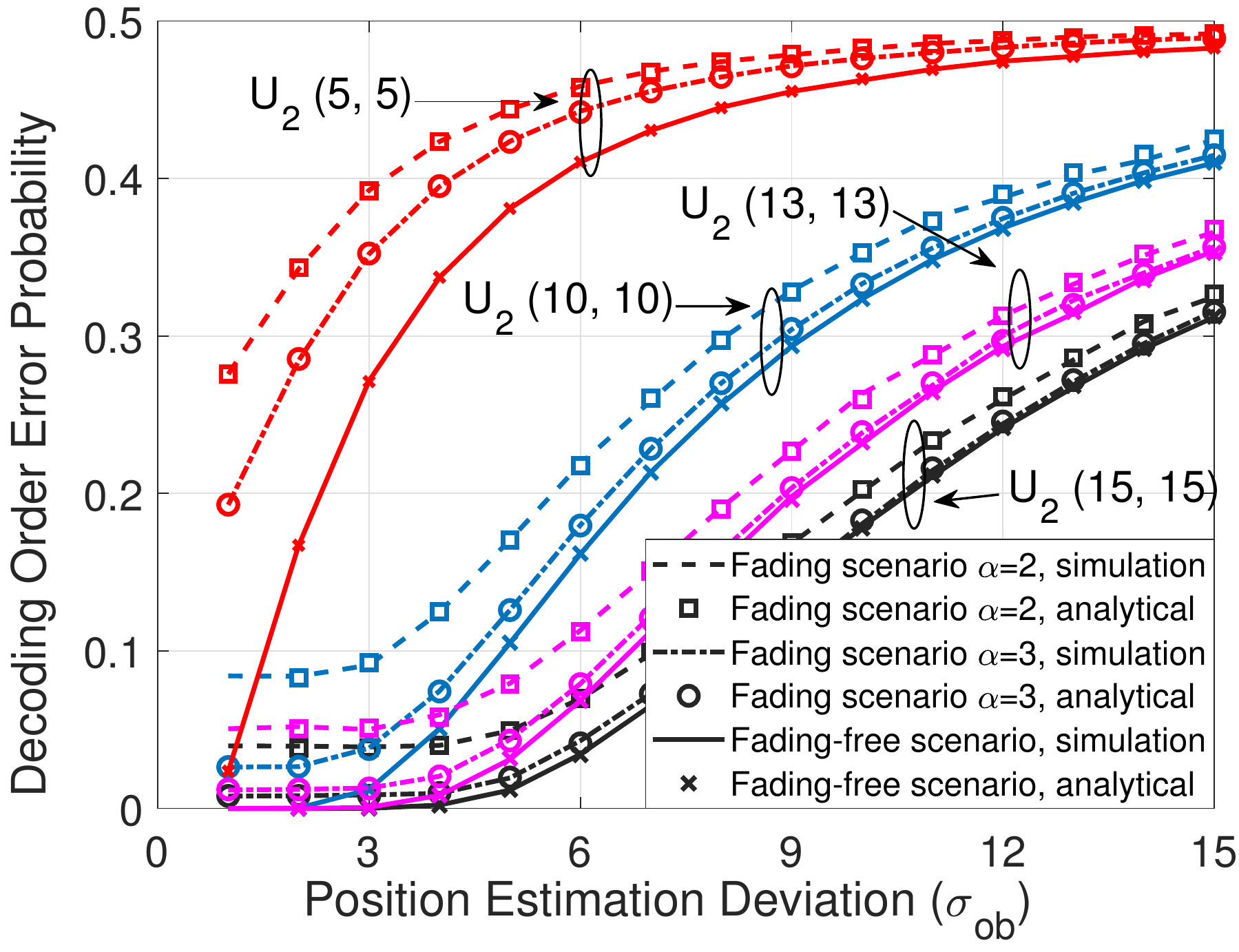}\vspace{-1em}
\caption{The decoding order error probability versus position estimation deviation with two users 
and fixed user position $U_1\left(3, 3\right)$. \vspace{-1em}}
\label{pe vs ob}
\end{figure}

In Fig. \ref{pe vs ob}, the impact of estimation deviation on 
the decoding order error probability $P_e$ is investigated.
The error probability increases as the position estimation error becomes larger. 
When users get closer, they are more likely to be disordered. 
For example, for Rayleigh fading with $\alpha=3$ and $\sigma_{ob}=3$,
the decoding order error probability is $0.35$ for $U_2$ position $\left(5, 5\right)$.
When $U_2$ is at the position $\left(10, 10\right)$, the error probability is $0.04$.
It is worthy to point out that the curves reveal distinct properties under different channel models. 
For the fading scenario, an error floor can still be observed even with very low deviation.
This is because incomplete channel information can lead to an incorrect decision on user decoding order.
As we can see, the error probability diminishes with larger path loss exponent $\alpha$,
which is in accordance with the results of \cite{RN500}. 
Therefore, the NOMA with partial channel information performs better in high path loss scenarios. 
The analytical result \eqref{pe awgn} matches with Monte Carlo simulation.

\begin{figure}[H]\centering
\includegraphics[width=0.45\textwidth]{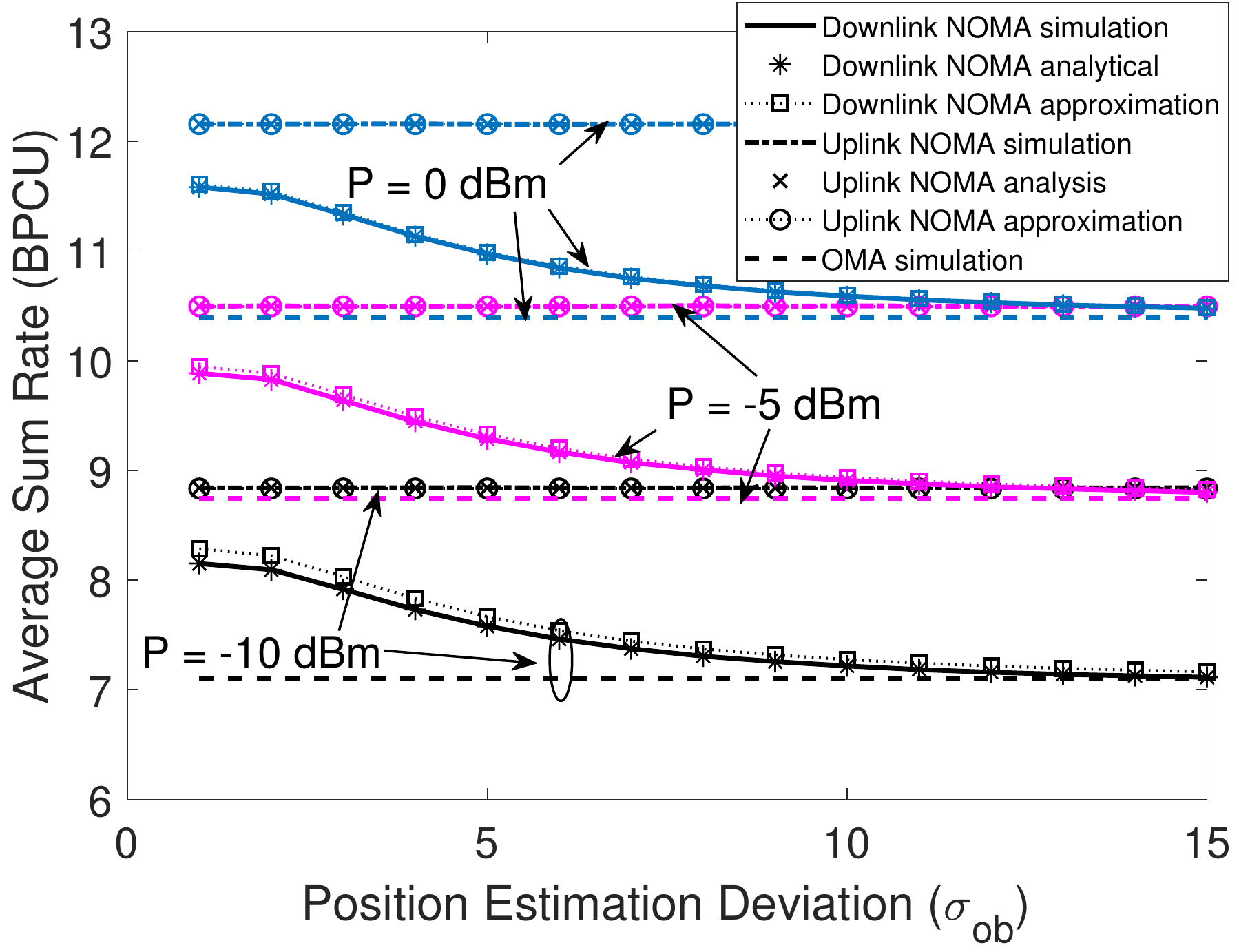}\vspace{-1em}
\caption{Impact of position observation noise on NOMA average sum rate (bit per channel use, BPCU) 
under different transmission power with $M=2$, $\alpha=2$, $\beta=0.8$, and 
fixed user position $U_1\left(3, 3\right)$, $U_2\left(7, 7\right)$. \vspace{-1em}}
\label{sr vs ob}
\end{figure}

In Fig. \ref{sr vs ob}, we demonstrate the influence of position estimation deviation 
on NOMA average sum rate and compare it with the OMA scheme.
As we can see, the average sum rate deteriorates 
when the observation noise variance $\sigma_{ob}^2$ increases.
This is because the decoding order error will reduce user rate and cause more detection failure.
In accordance to our previous analysis, 
the average sum rate of uplink is irrelevant to the user order. 
The comparison with OMA shows that the performance gain of NOMA can still be guaranteed 
when the estimation error is limited. 
However, NOMA becomes suboptimal when the positioning deviation is large, 
which shows the importance of accurate position information.

\begin{figure}[H]\centering
\includegraphics[width=0.45\textwidth]{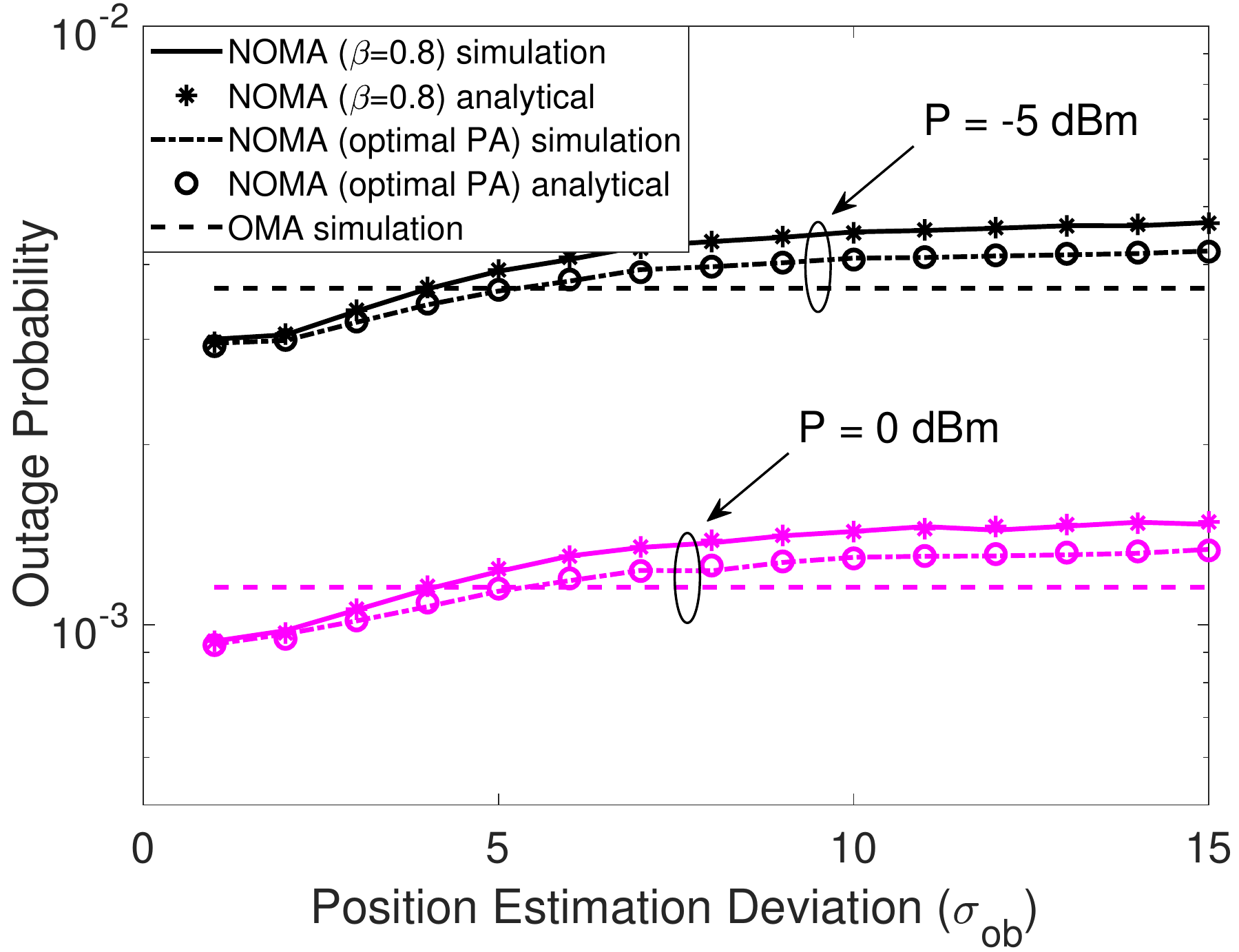}\vspace{-1em}
\caption{Impact of position observation noise on downlink outage probability 
under different transmission power with $M=2$, $\alpha=2$, $R_0^*=0.5$ bit per channel use (BPCU),
and fixed user position $U_1\left(3, 3\right), U_2\left(7, 7\right)$. \vspace{-1em}}
\label{down_cop vs ob}
\end{figure}

\begin{figure}[H]\centering
\includegraphics[width=0.45\textwidth]{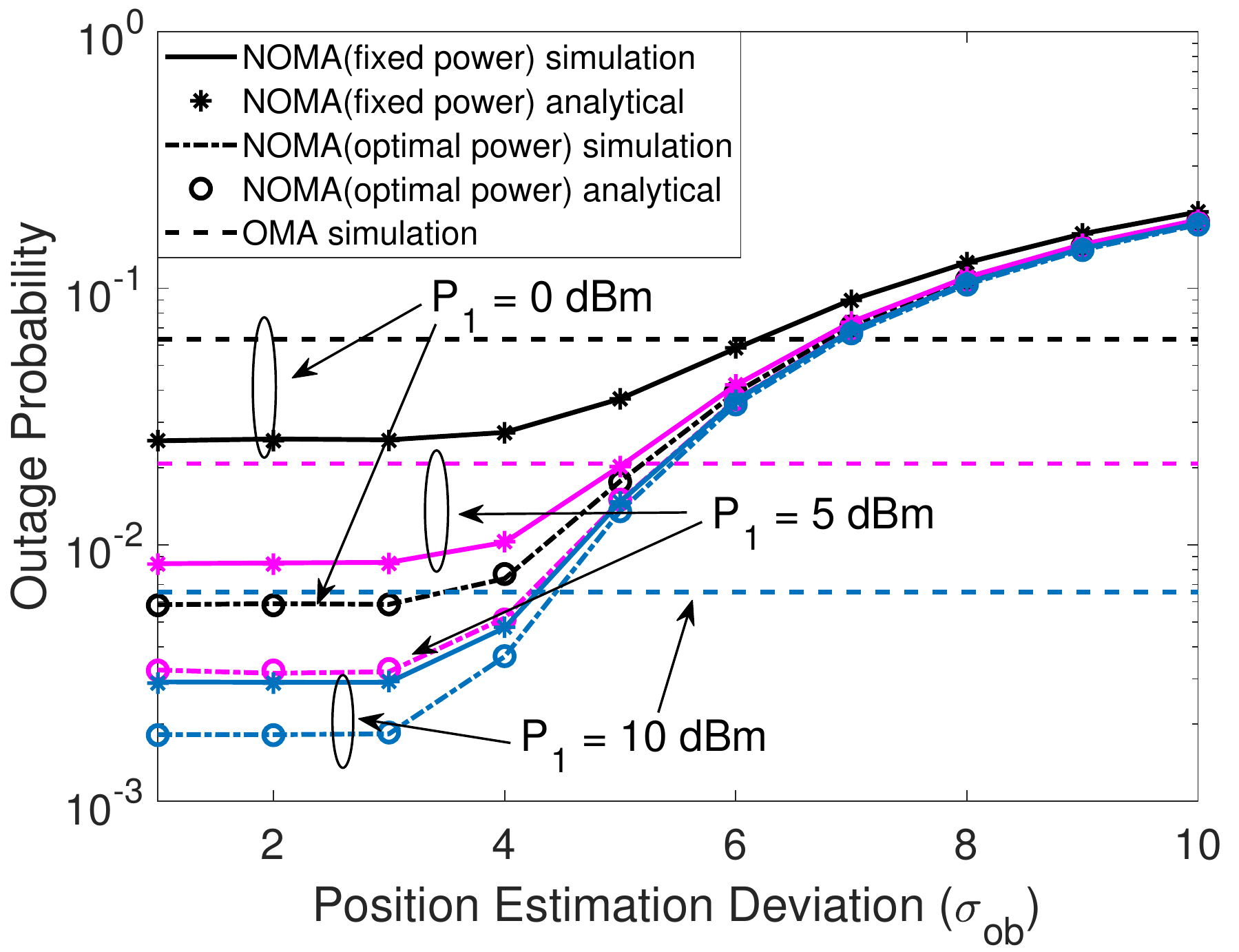}\vspace{-1em}
\caption{Impact of position observation noise on uplink outage probability 
under different transmission power with $M=2$, $\alpha=3.5$, $P_2=20dBm$, $R_0^*=0.1$ BPCU,
and fixed user position $U_1\left(3, 3\right), U_2\left(15, 15\right)$. \vspace{-1em}}
\label{up_cop vs ob}
\end{figure}

The outage performance for downlink and uplink NOMA transmission is shown 
in Fig. \ref{down_cop vs ob} and Fig. \ref{up_cop vs ob}. 
The numerical results are in accordance with the Monte Carlo simulation. 
The results have shown that the proposed dynamic signal power schemes improve
their outage performance. 
However, NOMA may still be worse than OMA when position deviation is high 
even if the optimal power strategy is adopted.
We can observe that compared to downlink, the outage probability of uplink scenario 
is more sensitive to position deviation.
This inspires us to utilize the hybrid NOMA/OMA scheme in uplink transmission. 
And it also indicates the importance of reducing position estimation error in order to further improve the outage performance.

\begin{table}[!t]\center
\caption{Simulation Parameters of Mobility Models}
\label{sim para}
\begin{tabular}{|p{150pt}<{\centering}|p{80pt}<{\centering}|}
\hline
\multicolumn{2}{|c|}{\textbf{Random Walk (RW)}} \\
\hline
Minimum Speed (m/s) & $0$ \\
\hline
Maximum Speed (m/s) & $2$ \\
\hline
Movement Interval (sample points) & 30 \\
\hline
$R_D$ & 30 \\
\hline
\multicolumn{2}{|c|}{\textbf{Random Waypoint (RWP)}} \\
\hline
Minimum Speed (m/s) & $1$ \\
\hline
Maximum Speed (m/s) & $3$ \\
\hline
Max Pause Time (sample points) & 5 \\
\hline
$R_D$ & 50 \\
\hline
\multicolumn{2}{|c|}{\textbf{Gauss-Markov Model (GM)}} \\
\hline
Speed Variance & 2 \\
\hline
Tuning Parameter & 0.5 \\
\hline
$R_D$ & 30 \\
\hline
\end{tabular}
\end{table}
 
\begin{table}[!t]\center
\caption{Comparison of Three Mobility Models}
\label{mobility model}
\begin{tabular}{|p{45pt}<{\centering}|p{45pt}<{\centering}|p{45pt}<{\centering}|p{45pt}<{\centering}|}
\hline
$\sigma_{ob}$ & RW & RWP & GM \\
\hline
$5$ & $2.45$ & $3.45$ & $2.53$ \\
\hline
$\sqrt{50}$ & $2.99$ & $4.26$ & $3.09$ \\
\hline
\end{tabular}
\end{table}

In order to evaluate the performance in mobile scenarios, three mobility models are considered, 
i.e., random walk (RW), random waypoint (RWP), and Gauss-Markov model (GM).
In random walk model, a user moves to a new location by randomly choosing a direction and speed. 
Users are allowed to move for an interval before changing their speed and direction.
In random waypoint model, users could stay at a location
in pause time and then choose the next destination. 
As for the Gauss-Markov model, the speed and direction are updated at each time slot. 
The tuning parameter is used to controls the level of randomness. 
The detailed model parameters are shown in Table \ref{sim para}.

Table \ref{mobility model} illustrates the effectiveness of position filtering
because the position deviation decreases for all the three cases,
which is beneficial to reduce decoding order error.
Taking the GM model as an example and recalling Fig. \ref{pe vs ob}, 
the decoding order error probability decreases rapidly
as $\sigma_{ob}$ changes from $5$ to $2.53$, especially when the two users get close.
To demonstrate the performance of our scheme, 
simulation results of spectrum efficiency and outage performance with GM mobility model 
are given in the following.

\begin{figure}[H]
\begin{center}
\subfigure[$M=2$]{\label{down_sr_u2 vs snr}\includegraphics[width=0.45\textwidth]{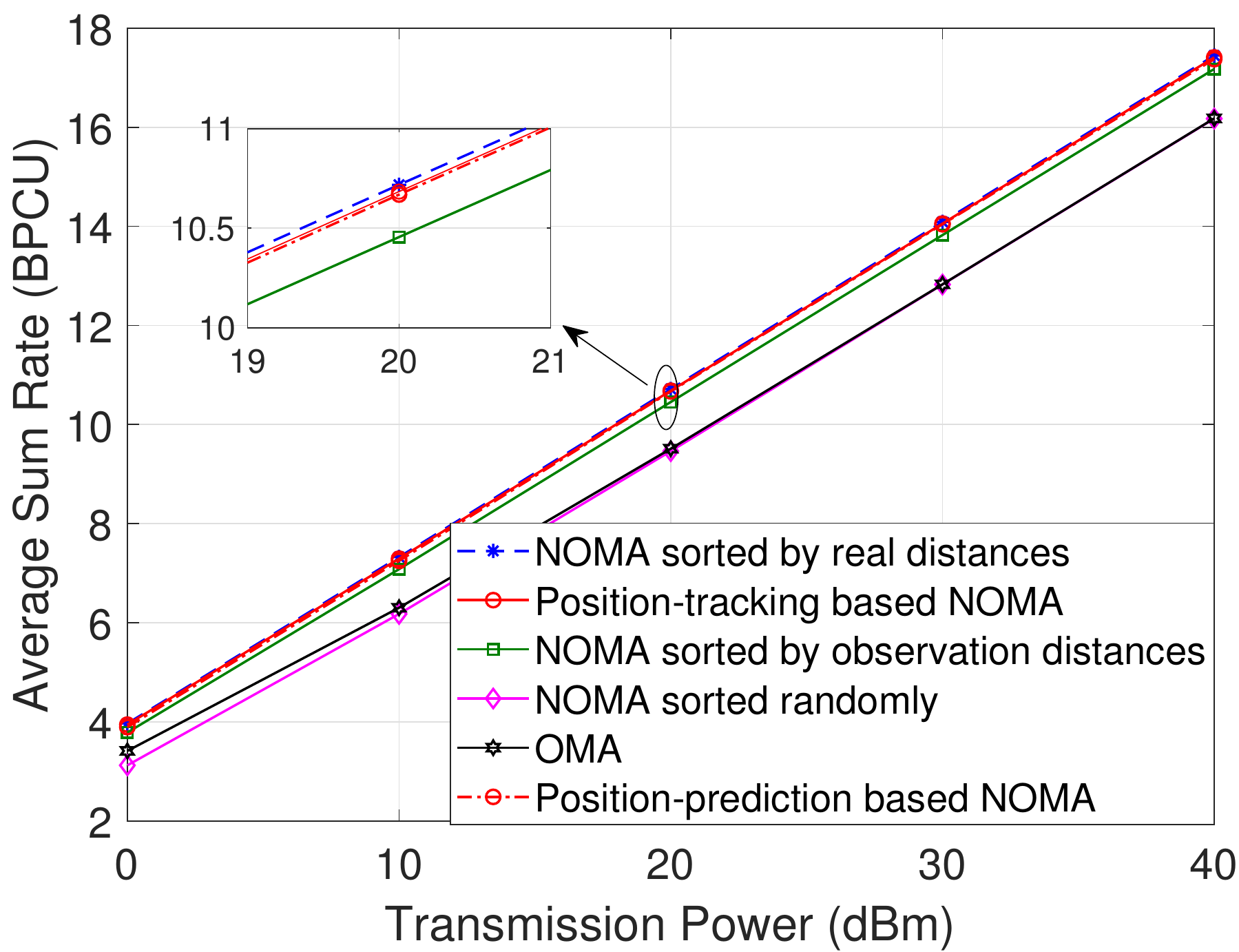}}
\subfigure[$M=5$, Pairing]{\label{down_sr_u5 vs snr}\includegraphics[width=0.45\textwidth]{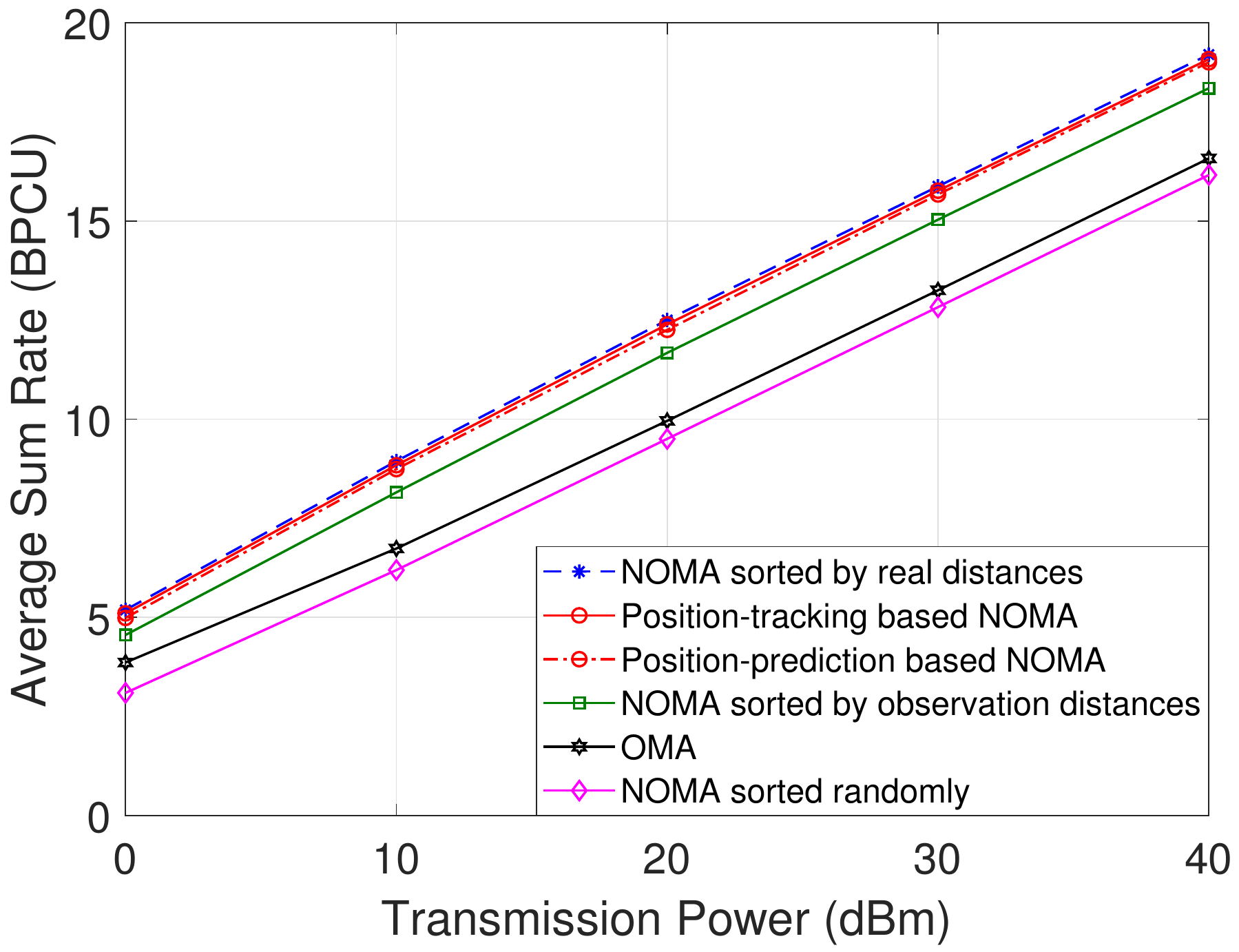}}
\vspace{-1em}
\end{center}
\caption{Average sum rate performance of downlink position filtering-based NOMA 
        with observation noise variance $\sigma_{ob}^2=50$ and $\beta=0.75$.}
\label{down_sr vs snr}
\end{figure}

In Fig. \ref{down_sr vs snr}, we demonstrate the simulation results 
for downlink average sum rate performance of our proposed NOMA schemes.
As we can see, the curves of the position filtering-based NOMA are close to that with 
perfect position knowledge, which is the upper bound of distance-based NOMA.
For position prediction-based NOMA, the slots in which the BS has access to observation data 
only account for $25\%$ in our simulation. 
But the algorithm with incomplete observation data will still obtain an excellent 
average sum rate performance. 
User pairing is considered in Fig. \ref{down_sr_u5 vs snr}, where the nearest user and farthest user 
are paired according to the estimated distances.
It reveals that the sum rate gap between real distances and observation distances 
becomes larger compared with Fig. \ref{down_sr_u2 vs snr} 
where two random users in the network can be paired.  
This property indicates that position estimation error has a greater influence on 
the user pairing-based NOMA.
Our scheme is more effective when a suitable pairing strategy is adopted
due to accurate position estimation.

\begin{figure}[H]
\begin{center}
\subfigure[$M=2$]{\label{down_cop_u2 vs snr}\includegraphics[width=0.45\textwidth]{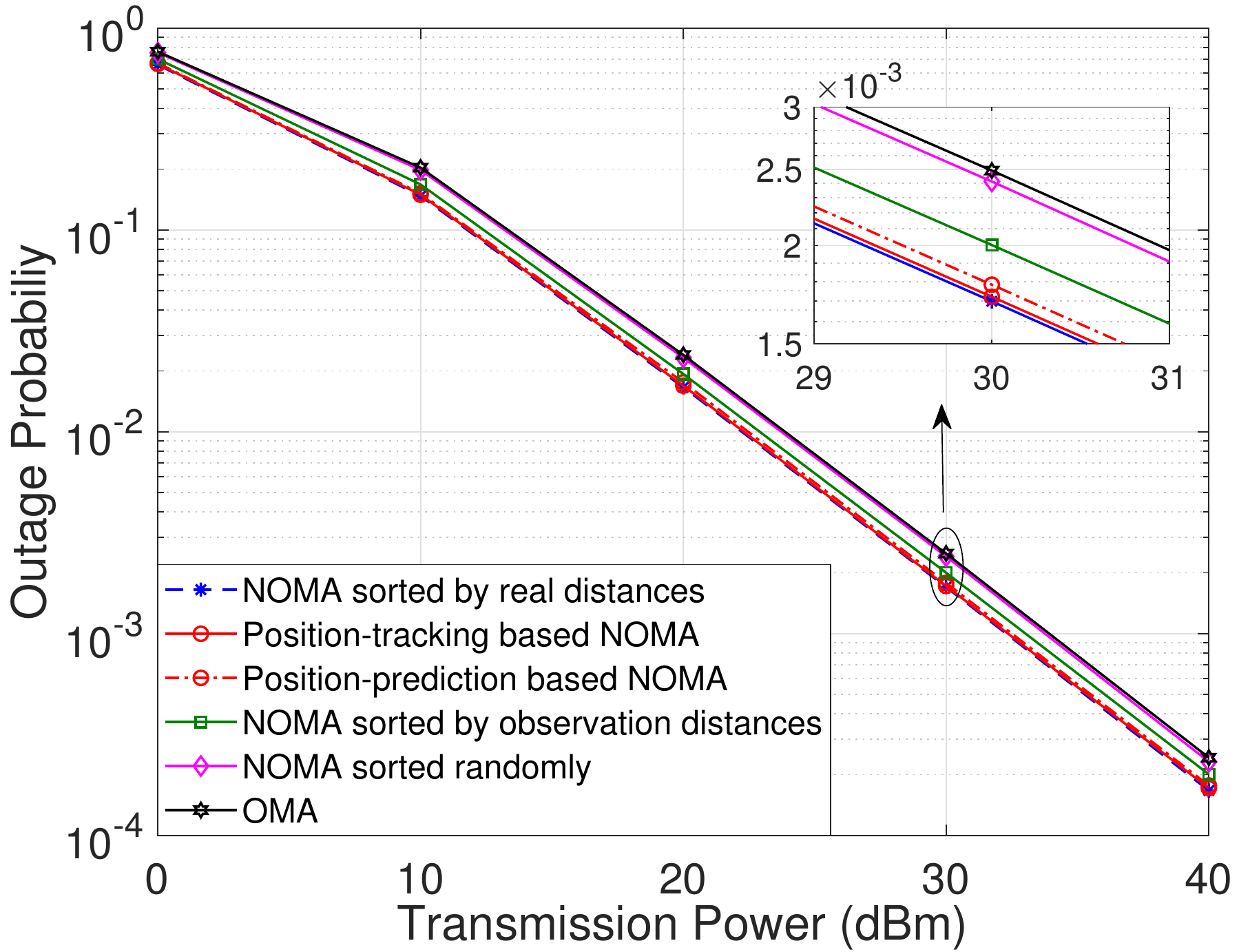}}
\subfigure[$M=5$, Pairing]{\label{down_cop_u5 vs snr}\includegraphics[width=0.45\textwidth]{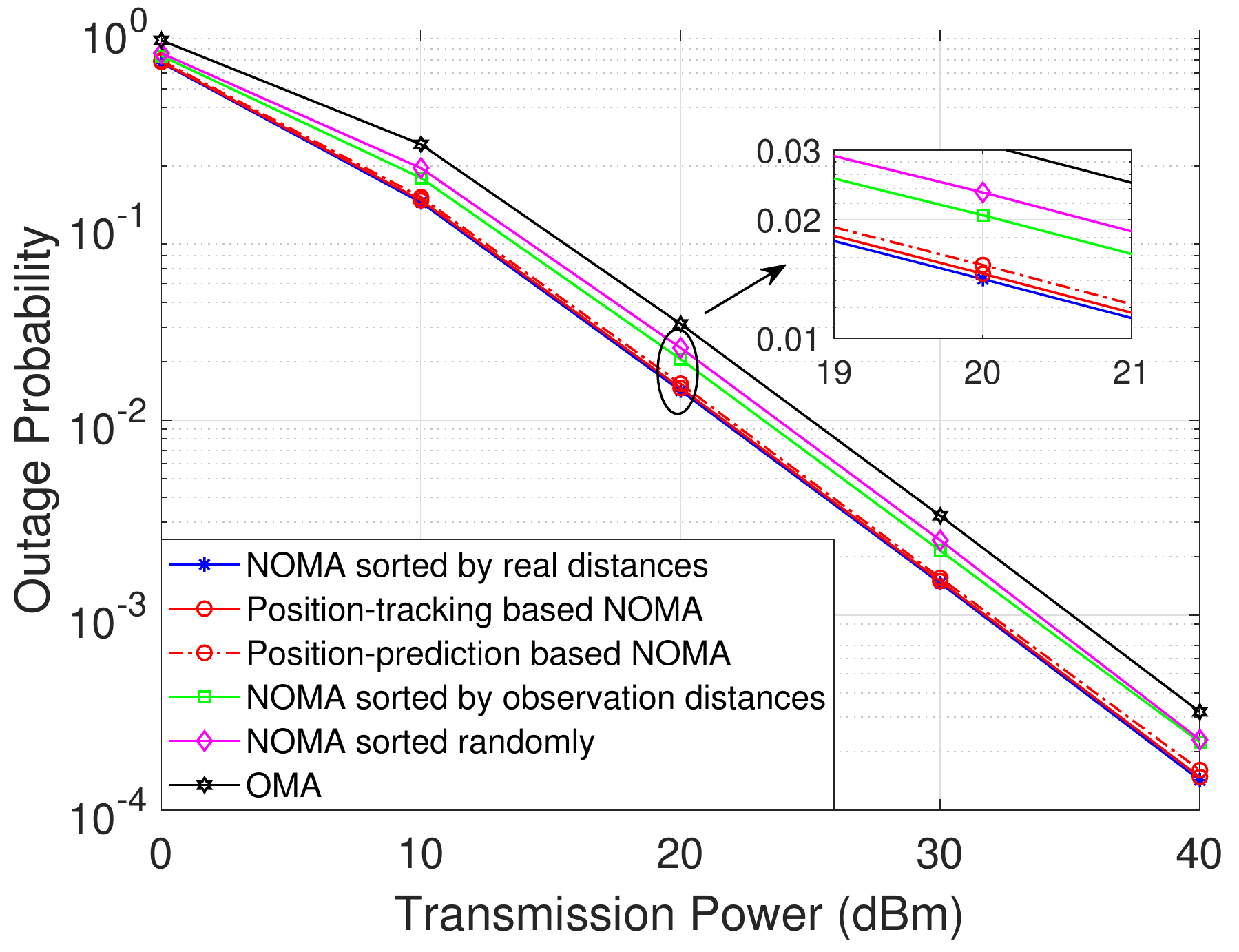}}
\vspace{-1em}
\end{center}
\caption{Outage performance of downlink position filtering-based NOMA with DPA,
        observation noise variance $\sigma_{ob}^2=50$, and target rate $R_0^*=1.5$ BPCU.}
\label{down_cop vs snr}
\end{figure}

Fig. \ref{down_cop vs snr} shows the outage performance simulation results
for both pair-based and not pair-based cases.
As shown from the figures, our proposed NOMA schemes approach the curves of real distance information. 
The outage probability of position prediction-based NOMA is almost the same as 
position tracking-based NOMA even if less position observation is used. 
As expected, the performance of our scheme is still superior to the OMA.

\begin{figure}[H]
\begin{center}
\subfigure[Outage Probability for User 1]{\label{down_d3_U1 vs tar}\includegraphics[width=0.45\textwidth]{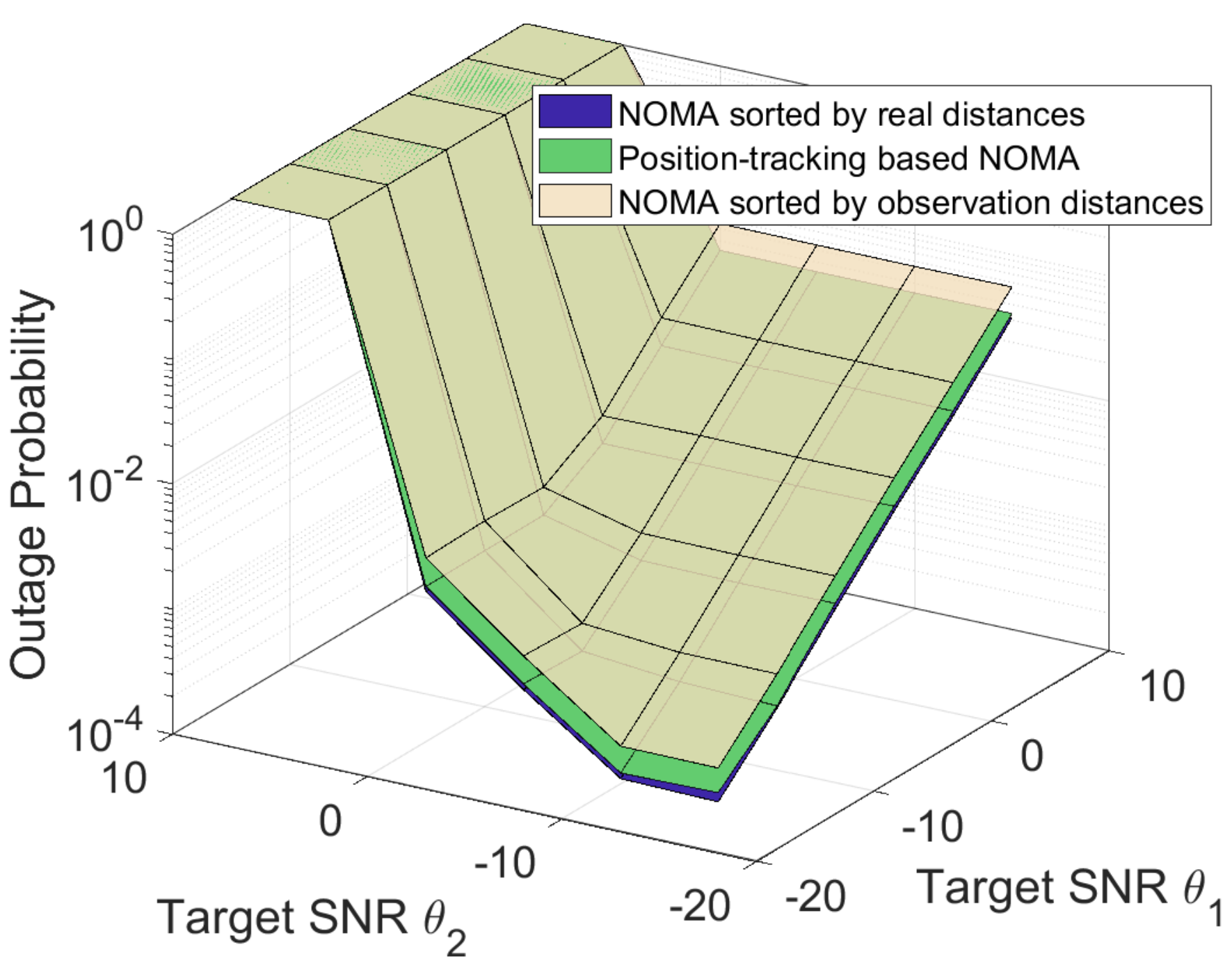}}
\subfigure[Outage Probability for User 2]{\label{down_d3_U2 vs tar}\includegraphics[width=0.45\textwidth]{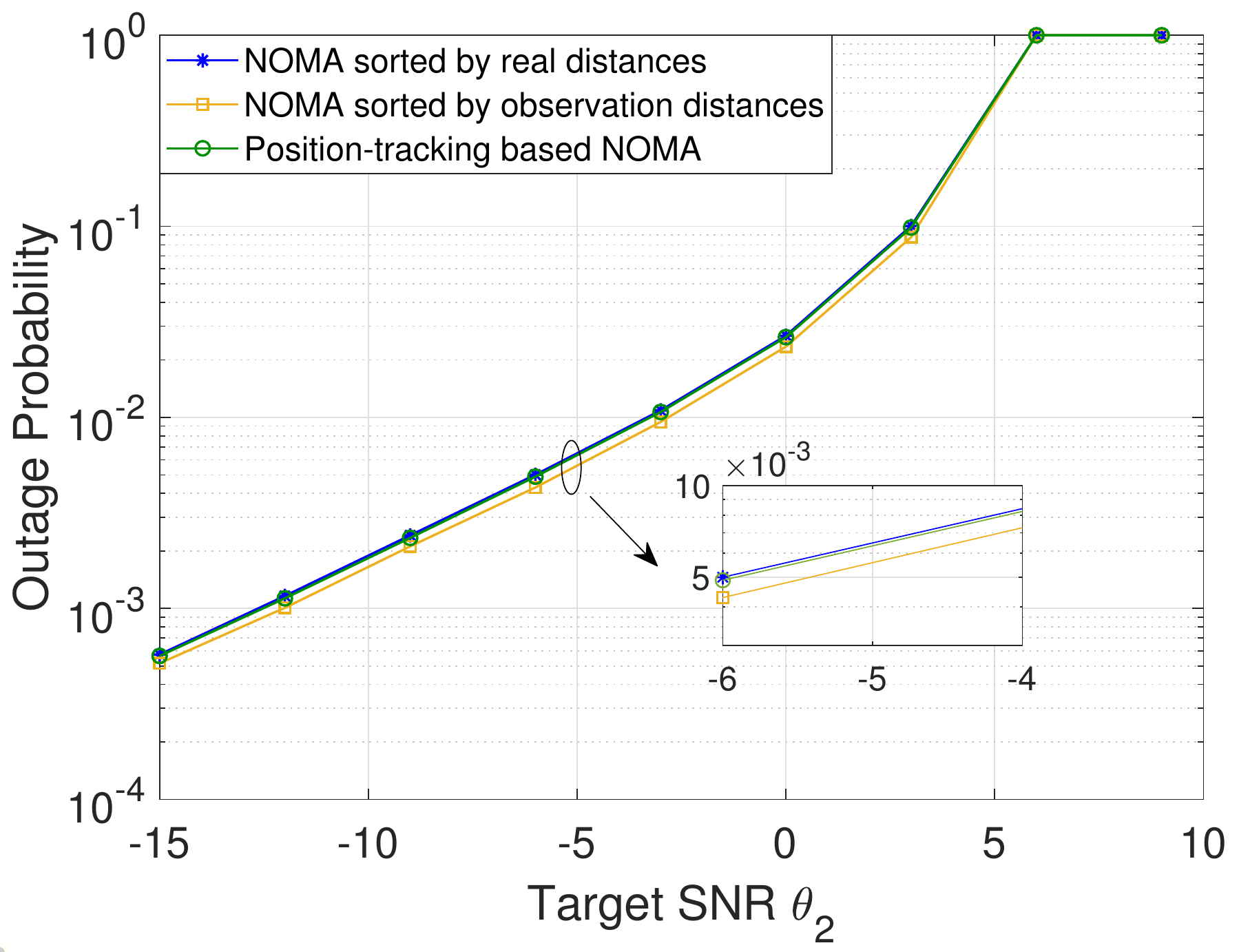}}
\vspace{-1em}
\end{center}
\caption{Outage performance for each user of downlink NOMA versus the target rate of 
        each signal with $\beta=0.75$, observation noise variance $\sigma_{ob}^2=50$ and 
        transmission power $P=15dBm$.}
\label{down_d3 vs tar}
\end{figure}

Fig. \ref{down_d3 vs tar} shows the relation between the outage performance of each user 
and target SNR of each signal in downlink transmission.
The target SNR $\theta_2$ affects the outage probability of both two users, whereas $\theta_1$ 
has nothing to do with the outage probability for user 1. 
It can be observed from Fig. \ref{down_d3_U1 vs tar} that the outage performance of 
position tracking-based NOMA is better than the NOMA sorted by observation distances. 
Actually, the improvement is at the cost of losing negligible outage performance 
for user 2 as shown in Fig. \ref{down_d3_U2 vs tar}. 
Furthermore, it is worth pointing out that the choice of target SNR $\theta_2$ 
also plays a key role in the NOMA with partial CSI. 
Unsuitable $\theta_2$ will make the outage probability of each user always be one 
due to the interference-limited property of $S_2$.

\begin{figure}[H]\centering
\includegraphics[width=0.45\textwidth]{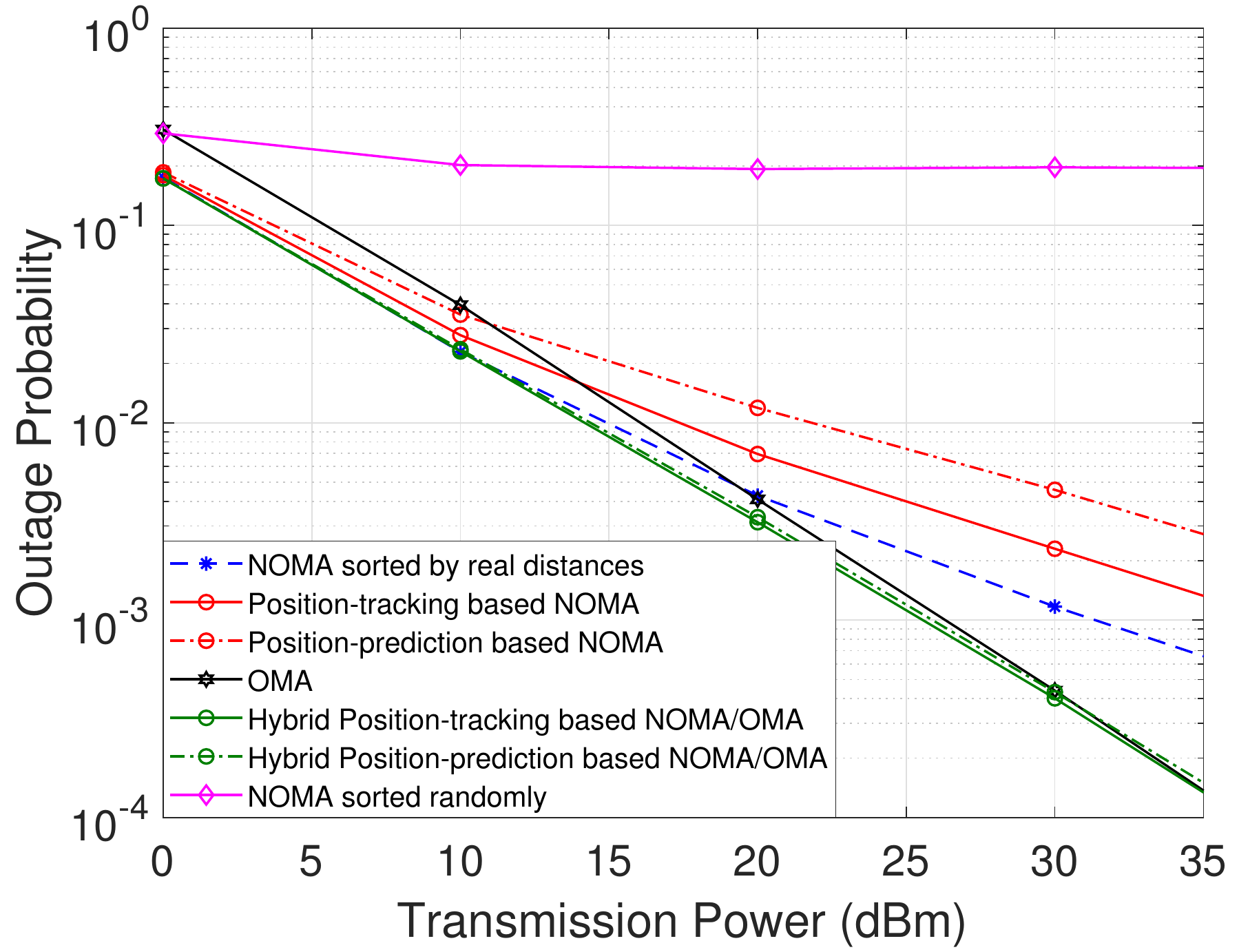}\vspace{-1em}
\caption{Outage performance of uplink position filtering-based NOMA with DPC,
observation noise variance $\sigma_{ob}^2=50$, $\alpha=3.5$, target rate $R_0^*=0.1$ BPCU,
and $M=5$ (pairing). \vspace{-1em}}
\label{up_cop vs snr}
\end{figure}

The outage performance of uplink NOMA is given in Fig. \ref{up_cop vs snr} 
when optimal power control strategy is adopted.
The error floor does not exist in our scheme.
One can be observed that position filtering-based NOMA is a better candidate 
in the relatively low SNR region, whereas OMA is superior when SNR is high.
Thus, to take advantages of the two schemes, a hybrid NOMA/OMA uplink scheme is developed 
for the practical system. 
The COP performance is derived firstly according to \eqref{up po eq}, \eqref{oma po},
and the better strategy is chosen adaptively for each uplink transmission. 
The results of hybrid position filtering-based NOMA/OMA scheme are illustrated in the figure. 
It can be observed that this hybrid scheme is superior to either single multiuser access protocol. 

\subsection{MIMO extension and discussions}
The SISO scenario is considered in the previous analysis. 
Actually, the NOMA scheme can also be combined with MIMO to achieve high spectrum efficiency. 
The existed MIMO-NOMA systems can usually be divided into two categories,
i.e., multi-cluster NOMA and single-cluster NOMA.

For the first case, each spatial freedom can support a cluster of users and 
spatial multiplexing is used to remove inter-cluster interference like \cite{RN365, zeng2019}.
Users in the same cluster perform NOMA the same way as SISO-NOMA does. 
In general, channel gains are highly dependent on beamforming and detection methods, 
thus the user order is usually determined by channel gain feedback.
Especially, when signal alignment is employed among the users in the same cluster\cite{ding2016}
or their channels are highly correlative, distance-based user order becomes a reasonable alternative.
In this case, the effect of user order is reduced to SISO-NOMA user order problem 
as analyzed previously.

For the second case, SU-MIMO is extended to multi-user scenarios by superposing signals 
with different power levels directly\cite{sun2015}. 
Generally, each user can occupy all the MIMO subchannels and each subchannel is shared 
by multiple users. 
In this case, users are usually ranked by distance 
because their channels are modeled by vectors (or matrix). 
Similar to the SISO scenario, our proposed position-filtering algorithms will reduce
the decoding order error probability due to more accurate position estimation.
However, the channel correlation needs to be considered in the MIMO scenarios.
The exact closed-form analysis is quite involved and so is the dynamic power scheme. 
Thus, this work is beyond the scope of this paper and is likely a promising direction. 

\section{conclusion}\label{s6}
In this paper, we have studied position information-based NOMA in mobile scenarios.
The impact of position deviation has been investigated for both downlink and uplink transmission.
Dynamic signal power schemes are proposed in order to further improve outage performance.
Our analysis has shown that the NOMA superiority over OMA will be reduced due to 
the position error.
Two position filtering-based algorithms have been proposed, i.e., position tracking-based NOMA and position prediction-based NOMA. 
By comparing our scheme to the conventional NOMA and OMA schemes, 
the improvement of average sum rate and outage performance is observed 
from the Monte Carlo simulation. 

\appendices

\section{Proof for Lemma \ref{LEMMA1}} \label{fading-free pe pr}
Denote the real position of $U_k$ by $\left(x_k,y_k\right)$ 
and the estimated position by $\left(\hat{x}_k,\hat{y}_k\right)$. 
Thus the real distance is $d_k^2=x_k^2+y_k^2$ and the distance estimation is 
$\hat{d}_k^2=\hat{x}_k^2+\hat{y}_k^2$.
Recall that the position deviation is subject to a Gaussian distribution, 
i.e. $\hat{x}_k\sim \mathcal{N}\left(x_k,\sigma_{ob}^2\right)$ and 
$\hat{y}_k\sim \mathcal{N}\left(y_k,\sigma_{ob}^2\right)$. 
Therefore, $\hat{d}_k^2$ follows a non-central chi-squared distribution. 
The probability density function (PDF) is given as follows:
\begin{align}\label{d pdf}
f_{\hat{d}_k^2}\left(x\right)=\frac1{2\sigma_{ob}^2}\exp\left(-\frac{x+\lambda_k}{2\sigma_{ob}^2}\right)
	I_0\left(\frac{\sqrt{x\lambda_k}}{\sigma_{ob}^2}\right), \left(x\ge 0\right),
\end{align}
where $\lambda_k=d_k^2$, $I_0\left(x\right)$ denotes zero-order modified Bessel function of the first kind. 
By expressing it in the form of a series and substituting into \eqref{d pdf}, we have
\begin{align}\nonumber
f_{\hat{d}_k^2}\left(x\right) &=\frac1{2\sigma_{ob}^2}\exp\left(-\frac{x+\lambda_k}{2\sigma_{ob}^2}\right)
	\sum_{i=0}^{\infty}\frac1{i!\Gamma\left(i+1\right)}
	\left(\frac{\sqrt{x\lambda_k}}{2\sigma_{ob}^2}\right)^{2i} \\
							  &= \sum_{i=0}^\infty P_{\lambda_k\beta}\left(i\right)
							  	f_{\alpha_i,\beta}\left(x\right),
\label{pdf series}
\end{align}
where $\alpha_i=i+1$, $\beta=\frac1{2\sigma_{ob}^2}$, 
$P_\lambda\left(k\right)=\frac{e^{-\lambda}\lambda^k}{k!}\left(k\ge 0\right)$ is the 
probability mass function (PMF) of a Poisson distribution, $f_{\alpha,\beta}\left(x\right)=
\frac{\beta^\alpha x^{\alpha-1}e^{-\beta x}}{\Gamma \left(\alpha\right)}, 
\left(x>0\right)$ is the PDF of Gamma distribution with parameters $\alpha$ and $\beta$. 
\eqref{pdf series} indicates that the PDF of $\hat{d}_k^2$ is a weighted sum of 
the Gamma distribution PDF, whose weight coefficients are the probabilities of Poisson distributions. 
Thus its the cumulative density function (CDF) is calculated naturally as
\begin{align}\label{cdf series}
F_{\hat{d}_k^2}\left(x\right)=\sum_{i=0}^\infty P_{\lambda_k\beta}\left(i\right)
	F_{\alpha_i,\beta}\left(x\right),
\end{align}
where $F_{\alpha,\beta}\left(x\right)=\frac{\gamma\left(\alpha,\beta x\right)}{\Gamma\left(\alpha\right)}$ 
is the CDF of Gamma distribution and is expressed as the form of a regularized Gamma function. 

In fading-free scenarios, user channel gains are only determined by their path loss. 
Recall that $d_1<d_2$, we have $\absS{r_1}>\absS{r_2}$.
According to \eqref{pe def}, the decoding order error event occurs only if the order of distance estimation
is incorrect, i.e. $\left\{\hat{d}_1>\hat{d}_2\right\}$. 
The error probability is formulated as
\begin{align}\label{pe ori}
P_e^1=& \prb{\hat{d}_1>\hat{d}_2}= \prb{\hat{d}_1^2>\hat{d}_2^2} \nonumber \\
	=&\int_0^{+\infty}\int_0^uf_{\hat{d}_1^2}\left(u\right)f_{\hat{d}_2^2}\left(v\right)dvdu.
\end{align}

Substitute \eqref{pdf series} to \eqref{pe ori}, a double integral is obtained.
The inner integral result is the CDF of Gamma distribution as shown in \eqref{cdf series}. 
After simplification, we obtain the expression shown in \eqref{pe awgn}. 
The integral term $I_{i,j}$ is calculated as follows:
\begin{align}
&I_{i,j} =\int_0^\infty f_{\alpha_i,\beta}\left(x\right)F_{\alpha_j,\beta}\left(x\right)dx \nonumber \\ 
\overset{\left(a\right)}{=}& \left(\frac12\right)^{\alpha_i+\alpha_j}\binom{\alpha_i+\alpha_j-1}{\alpha_j}
	F\left(1,\alpha_i+\alpha_j,\alpha_j+1,\frac12\right),
\label{Iij detail}
\end{align}
where the calculation process $\left(a\right)$ in \eqref{Iij detail} is obtained 
by using the result (6.455) in \cite{RN416} and $F\left(\cdot\right)$ denotes a hypergeometric function. 
Thus, the proof is completed.

\section{Proof for Proposition \ref{PROSR1}} \label{down sr pr}
Recall that the distances of the two users satisfy $d_1<d_2$ and
the results of distance estimation can lead to a different power allocation and detection process. 
To make it clearer, we show the decoding process and the signals that restrict the achievable rate 
(marked as boldface) in Table \ref{chart ray}.

\begin{table}[!t]
\caption{Receiver Detection Process of Rayleigh Channels}
\begin{tabularx}{\linewidth}{>{\setlength{\hsize}{1.4\hsize}}Y >{\setlength{\hsize}{0.8\hsize}}Y  >{\setlength{\hsize}{0.8\hsize}}Y}
\toprule
& $U_1$ & $U_2$ \\ 
\midrule
$\left|r_1\right|^2>\left|r_2\right|^2$, $\hat{d}_1>\hat{d}_2$ & $S_1$ & $\bm{S}_1\rightarrow\bm{S}_2$ \\
$\left|r_1\right|^2<\left|r_2\right|^2$, $\hat{d}_1<\hat{d}_2$ & $\bm{S}_2\rightarrow\bm{S}_1$ & $S_2$ \\
$\left|r_1\right|^2>\left|r_2\right|^2$, $\hat{d}_1<\hat{d}_2$ & $S_2\rightarrow\bm{S}_1$ & $\bm{S}_2$ \\
$\left|r_1\right|^2<\left|r_2\right|^2$, $\hat{d}_1>\hat{d}_2$ & $\bm{S}_1$ & $S_1\rightarrow\bm{S}_2$ \\
\bottomrule
\end{tabularx}
\label{chart ray}
\end{table}

As shown in the table, the detection process is divided into four cases 
according to distance estimation and the amplitude of fading. Take the first case for example. 
User $U_1$ regards the signal $S_2$ as interference to detect its own message directly. 
The user $U_2$ firstly considers its own signal $S_2$ as interference to detect $S_1$ 
and removes $S_1$ according to SIC strategy, then detects the signal $S_2$. 
So as we can see, the achievable rate of both $S_1$ and $S_2$ are limited by $U_2$ 
because of its poor channel condition. 
Let $P_k\left(k=1,\ldots,4\right)$ denotes the probability of each case. 
As a result, the average sum rate is expressed as 
\begin{align}\label{R ray ori}
R_{sum}^{\Rmnum{1}}=\sum_{k=1}^4\E\left\{C_k\right\}P_k,
\end{align}
where 
\begin{align}
C_1 =& \log\left(1+\frac{\left|r_2\right|^2\beta}{\left|r_2\right|^2\left(1-\beta\right)+\frac1\rho}\right) \nonumber \\
     &+\log\left(1+\rho\left|r_2\right|^2\left(1-\beta\right)\right),
\end{align}
and other $C_k$ can be expressed similarly. The key to obtain the average sum rate 
$R_{sum}^{\Rmnum{1}}$ in \eqref{R ray ori} is to calculate the conditional expectation of each case. 
Let random variable $X_k=\left|r_k\right|^2\left(k=1,2\right)$, and the first expectation term 
is calculated as follows:
\begin{align}\nonumber
 &\E\left\{C_1|X_1>X_2,\hat{d}_1>\hat{d}_2\right\} \\ \nonumber
=&\int_0^\infty\int_{x_2}^\infty C_1\left(x_2\right)f\left(x_1,x_2|X_1>X_2\right)dx_1dx_2 \\
=& \left.{\frac{\lambda_2\rho}{\lambda\ln2}\int_0^\infty \frac{\e^{-\lambda x_2}}{1+\rho x_2}dx_2}
	\middle/{\pr\left\{X_1>X_2\right\}}\right.. \label{con expect 1}
\end{align}
Recall the following equation 
\begin{align}\label{int res}
\int_0^\infty\frac{\e^{-lx}}{1+x\phi}dx=-\frac{1}{\phi}\e^{\frac l\phi}\ei\left({-\frac l\phi}\right).
\end{align} 
Define
\begin{align}
\varphi\left(k,\phi\right)=&-\frac{\lambda_k}{\lambda \ln2}\e^{\frac\lambda\phi}
	{\mathrm{Ei}}\left(-\frac\lambda\phi\right), \label{fai1 def} \\
\varphi'\left(k,\phi\right)=&-\frac1{\ln2}\e^{\frac{\lambda_k}{\phi}}
	\ei\left(-\frac{\lambda_k}\phi\right). \label{fai2 def}
\end{align}
By substituting \eqref{pr fading}, \eqref{int res} and \eqref{fai1 def} into \eqref{con expect 1}, we have
\begin{align}\label{con expect 2}
\E\left\{C_1|X_1>X_2,\hat{d}_1>\hat{d}_2\right\}=\varphi\left(2,\rho\right)/\left(\frac{D}{D+1}\right).
\end{align} 

Other terms can be achieved with the similar method. Like \eqref{pe2 ori} and \eqref{pr fading}, 
the probability $P_k$ is obtained. 
By substituting the results into \eqref{R ray ori}, the first part is proved.

When $x<0$ and $x\rightarrow0$ , the exponential integral can be approximated as
\begin{align} 
\EI{x}\approx \LN{-x}+ C,
\end{align}
where $C$ represents the Euler's constant.
At high SNRs, i.e., $\phi\rightarrow\infty$, 
\eqref{fai1 def} and \eqref{fai2 def} are approximated by
\begin{align}
\fai{k}{\phi}\approx -\frac{\lambda_k}{\lambda\ln2}\left(C+\ln\frac{\lambda}{\phi}\right),
	\label{fai1 approx} \\
\faip{k}{\phi}\approx -\frac{1}{\ln2}\left(C+\ln\frac{\lambda_k}{\phi}\right).
	\label{fai2 approx}
\end{align}
After substituting \eqref{fai1 approx} and \eqref{fai2 approx} into \eqref{down sr eq},
the proof is completed.

\section{Proof for Proposition \ref{PROPO1}} \label{down po pr}
According to Appendix \ref{down sr pr}, the decoding process consists of four cases. 
Similar to the previous method, we need to analyze each case to formulate the outage performance.
The COP can be calculate as 
$P_{cop}^{\Rmnum{1}}=\sum_{k=1}^{4}P_{cop}^kP_k \label{down cop}$.
In accordance to Table \ref{chart ray}, the outage probability of case 1, i.e., $P^1_{cop}$, 
is expressed like this:
\begin{align}
P^1_{cop} =& 1-\mathrm{Pr}\left\{\frac{X_2\beta}{X_2\left(1-\beta\right)+\frac1\rho}>\epsilon_0,
	\rho X_2\left(1-\beta\right)>\epsilon_0\right. \nonumber \\
		& \left.\bigg|X_1>X_2,\hat{d}_1>\hat{d}_2\right\}. \label{down p1 ori}
\end{align}

Note that the signal $S_1$ needs to be decoded by both two users. An outage event will occur 
to $U_1$ inevitably if $U_2$ cannot detect $S_1$ successfully.
The equation \eqref{down p1 ori} is obtained on such a condition. After some mathematical manipulation, 
we have
\begin{align}
P^1_{cop} =& 1 - \frac{\prb{X_1>X_2>\zeta}}{\prb{X_1>X_2}}. \label{down p1 sim}
\end{align}

It is noteworthy that equation \eqref{down p1 sim} is conditioned on 
$\beta-\left(1-\beta\right)\epsilon_0>0$. Otherwise, the outage probability is always equal to one 
since $S_1$ can never be decoded successfully. 
The probability of the numerator is calculated by a double integral 
and the probability of the denominator has been derived in appendix \ref{down sr pr}. 
Thus the COP of case 1 is formulated as follows:
\begin{align}
P^1_{cop}=1-\frac{\lambda_2}{\lambda}\e^{-\lambda\zeta}/\left(\frac{D}{D+1}\right). \label{down p1 res}
\end{align}

Similarly, the outage probabilities of other cases can be calculated.
By substituting the results, the proof is completed.

\section{Proof for Corollary \ref{OPTPA}} \label{problemDown pr}
We can see from Proposition \ref{PROPO1} that when 
$\frac{\epsilon_0}{\epsilon_0+1}<\beta<\frac{\epsilon_0+1}{\epsilon_0+2}$, 
$P_{cop}^{\Rmnum{1}}=1-\e^{\frac{\lambda\epsilon_0}{\rho\left[\beta-\left(1-\beta\right)\epsilon_0\right]}}$. 
An observation is that the outrage probability monotonically decreases with $\beta$.

As for the region $\frac{\epsilon_0+1}{\epsilon_0+2}<\beta<1$, 
define the function $f\left(\beta\right)=\lambda_2A+\lambda_1B$. 
Thus, the monotonicity of $P_{cop}^{\Rmnum{1}}$ is determined by $f\left(\beta\right)$. 
The derivative of function with respect to $\beta$ is calculated as 
\begin{align} \label{f_dev}
\frac{df}{d\beta}=\frac{\epsilon_0\left(a\beta^2+b\beta+c\right)}
	{\rho\left(1-\beta\right)^2\left(\beta-\epsilon_0+\beta\epsilon_0\right)^2},
\end{align}
where $a=\left(1+\epsilon_0\right)\left[\lambda_1\left(1+\epsilon_0\right)-\lambda_2\right]$, 
$c=\epsilon_0^2\lambda_1-\lambda_2-\epsilon_0\lambda_2$,
and $b=2\left(1+\epsilon_0\right)\left(\lambda_2-\epsilon_0\lambda_1\right)$.
The zero points of \eqref{f_dev} exist and are given as follows:
\begin{align}
\beta^{\dagger}=\frac
	{\sqrt{1+\epsilon_0}\left(\epsilon_0\lambda_1-\lambda_2\right)\pm\sqrt{\lambda_1\lambda_2}}
	{\sqrt{1+\epsilon_0}\left[\lambda_1\left(1+\epsilon_0\right)-\lambda_2\right]}.
\end{align}

Three important features of the roots are shown in the following.
\begin{enumerate}
	\item The root with plus sign $\beta^+$ satisfies $\frac{\epsilon_0+1}{\epsilon_0+2}<\beta^+<1$. 
	Equivalently, the inequality is expressed as
	\begin{align}
	0 < \frac{\lambda_1\sqrt{1+\epsilon_0}-\sqrt{\lambda_1\lambda_2}}
		{\sqrt{1+\epsilon_0}\left[\lambda_1\left(1+\epsilon_0\right)-\lambda_2\right]} 
		< \frac{1}{\epsilon_0+2}.
	\end{align}

	Namely,
	\begin{align} \label{inq2}
	0 < \frac{\lambda_1}{\lambda_1\left(1+\epsilon_0\right)+
		\sqrt{\lambda_1\lambda_2\left(1+\epsilon_0\right)}}
		< \frac{1}{\epsilon_0+2}.
	\end{align}
	Recall that $\lambda_1<\lambda_2$ and all the variables are positive, 
	the inequality \eqref{inq2} is easily proven. 

	\item On the condition of $a>0$, i.e., $\lambda_1\left(1+\epsilon_0\right)>\lambda_2$, 
	the root with minus sign satisfies $\beta^-<\frac{\epsilon_0+1}{\epsilon_0+2}$. 
	The result can be proven with similar method above.

	\item In the case of $a<0$, $\beta^->1$, which is equivalent to 
	\begin{align} \label{inq3}
	\frac
		{\lambda_1\sqrt{1+\epsilon_0}+\sqrt{\lambda_1\lambda_2}}
		{\sqrt{1+\epsilon_0}\left[\lambda_1\left(1+\epsilon_0\right)-\lambda_2\right]}
		< 0.
	\end{align}

	Note that the numerator is positive and the denominator is negative. 
	Thus its proof has finished.
\end{enumerate}

To sum up, the COP is monotonically decreasing in the region $\left(\frac{\epsilon_0+1}{\epsilon_0+2},\beta^+\right)$ and monotonically increasing in $\left(\beta^+,1\right)$ in either case. Therefore, the proof is completed.

\section{Proof for Proposition \ref{PROPO2}} \label{up po pr}
When $\hat{d}_1<\hat{d}_2$, the BS decodes $S_1$ first and $S_2$ follows. 
The outage probability at the base station is formulated as
\begin{align}
P_{cop}^1 	=& 1 - \prb{\frac{\beta X_1}{\left(1-\beta\right)X_2+\frac 1\rho}>\epsilon_0,
	\rho\left(1-\beta\right)X_2>\epsilon_0} \nonumber \\
			=& 1 - \prb{X_1>kX_2+C, X_2>B}, \label{up p1 ori}
\end{align}
where $B=\frac{\epsilon_0}{\rho_2}$, $C=\frac{\epsilon_0}{\rho_1}$ 
and $k=\frac{\epsilon_0\rho_2}{\rho_1}$. 
The probability in \eqref{up p1 ori} is calculated by the integral of the joint PDF function 
over the region, which is illustrated as follows:
\begin{align}
p_{cop}^1 	=& 1 - \int_{B}^{\infty}\int_{kX_2+C}^{\infty}
	\lambda_1\e^{-\lambda_1X_1}\lambda_2\e^{\lambda_2X_2}dX_1dX_2 \nonumber \\
			=& 1 - \frac{\lambda_2}{\lambda_2+k\lambda_1}\e^{-\lambda_1C
				-\left(\lambda_2+k\lambda_1\right)B}. \label{up p1 final}
\end{align}

Similarly, the outage probability of the case $\hat{d}_1>\hat{d}_2$ can be expressed as follows:
\begin{align}
P_{cop}^2	= 1 - \frac{\lambda_1}{\lambda_1+k\lambda_2}\e^{-\lambda_2C
	-\left(\lambda_1+k\lambda_2\right)B}. \label{up p2 final}
\end{align}

Finally, add the weighted results \eqref{up p1 final} and \eqref{up p2 final}. 
The result of \eqref{up po eq} is obtained.

\section{Proof for Corollary \ref{OPTPC}} \label{problemUp pr}
The transmission power for both users is to be solved. 
To this end, the problem is separated into two steps,
i.e, the transmission SNR of user 1 is firstly optimized for a given $\rho_2$. 
After that, the optimal $\rho_2$ is calculated for the derived $\rho_1$. 

Recall that the outage is expressed as
\begin{align}
P_{cop}^{\Rmnum{2}} = 1 -\frac{\lambda_2}{\lambda_2+\frac{\epsilon_0\lambda_1\rho_2}
	{\rho_1}}\e^{-\frac{\lambda_1\left(\epsilon_0+\epsilon_0^2\right)}{\rho_1}
	-\frac{\lambda_2\epsilon_0}{\rho_2}}.
\end{align}
Note that the COP is minimized when $U_1$ transmits messages at its largest power, 
which is because a larger transmission power means less outage 
when the power of the interference signal is constant. 
Thus, the first part of the proof has finished. 
By deriving $P_{cop}^{\Rmnum{2}}$ with respect to $\rho_2$, we have
\begin{align} \label{cop_dev}
\frac{dP_{cop}^{\Rmnum{2}}}{d\rho_2} =& \frac
	{\Omega_1\epsilon_0\lambda_2\e^{-\frac{\lambda_1\left(\epsilon_0+\epsilon_0^2\right)}{\Omega_1}
	-\frac{\lambda_2\epsilon_0}{\rho_2}}}
	{\rho_2^2\left(\rho_2\epsilon_0\lambda_1+\Omega_1\lambda_2\right)^2} \nonumber \\
& \times \left(\lambda_1\rho_2^2-\epsilon_0\lambda_1\lambda_2\rho_2-\Omega_1\lambda_2^2\right).
\end{align}
An useful observation is that the fraction in \eqref{cop_dev} is positive. 
Therefore, the zero points are given by
\begin{align}\label{zp2}
\rho_2^{\dagger} = \frac{\epsilon_0\lambda_1\lambda_2\pm\lambda_2
	\sqrt{4\Omega_1\lambda_1+\epsilon_0^2\lambda_1^2}}{2\lambda_1}.
\end{align}

From \eqref{zp2}, it is easy to prove that $\rho_2^+>0$ and $\rho_2^-<0$. 
According to the monotonicity similar to Appendix \ref{problemDown pr}, 
the optimal transmission SNR for $U_2$ is obtained.
\begin{align}
\rho_2^* = \min\left(\Omega_2, \rho_2^+\right).
\end{align}

\end{document}